\documentclass[12pt,preprint]{aastex}
\shorttitle{}
\shortauthors{Nesvorn\'y et al.}
\begin{document}
\title{Jumping Neptune Can Explain the Kuiper Belt Kernel}
\author{David Nesvorn\'y}
%, David Vokrouhlick\'y$^2$, Alessandro Morbidelli$^3$}
\affil{Department of Space Studies, Southwest Research Institute, 1050 Walnut St., \\Suite 300, 
Boulder, CO 80302, USA} 
%\affil{(2) Institute of Astronomy, Charles University, V Hole\v{s}ovi\v{c}k\'ach 2, \\
%180 00 Prague 8, Czech Republic}
%\affil{(3) D\'epartement Cassiop\'ee, University of Nice, CNRS, Observatoire de la C\^ote d'Azur, \\Nice, 
%06304, France}

\begin{abstract}
The Kuiper belt is a population of icy bodies beyond the orbit of Neptune. A particularly puzzling and 
up-to-now unexplained feature of the Kuiper belt is the so-called `kernel', a concentration of orbits 
with semimajor axes $a\simeq44$~AU, eccentricities $e\sim0.05$, and inclinations $i<5^\circ$. Here we 
show that the Kuiper belt kernel can be explained if Neptune's otherwise smooth migration was 
interrupted by a discontinuous change of Neptune's semimajor axis when Neptune reached $\simeq 28$ AU. 
Before the discontinuity happened, planetesimals located at $\sim$40 AU were swept into Neptune's 
2:1 resonance, and were carried with the migrating resonance outwards. The 2:1 resonance was at $\simeq$44 AU 
when Neptune reached $\simeq28$ AU. If Neptune's semimajor axis changed by fraction of AU at this point, 
perhaps because Neptune was scattered off of another planet, the 2:1 population would have been released 
at $\simeq$44 AU, and would remain there to this day. We show that the orbital distribution of bodies 
produced in this model provides a good match to the orbital properties of the kernel. If Neptune migration 
was conveniently slow after the jump, the sweeping 2:1 resonance would deplete the population of bodies
at $\simeq$45-47 AU, thus contributing to the paucity of the low-inclination orbits in this region. 
Special provisions, probably related to inefficiencies in the accretional growth of sizable objects, are 
still needed to explain why only a few low-inclination bodies have been so far detected beyond 
$\simeq$47~AU. 
\end{abstract}
\section{Introduction}
Following the pioneering work of Malhotra (1993, 1995), studies of Kuiper belt dynamics first considered 
the effects of outward migration of Neptune that can explain the prominent populations of Kuiper Belt Objects
(KBOs) in major resonances (Hahn \& Malhotra 1999, 2005; Chiang \& Jordan 2002; Chiang et al. 2003; Levison \& 
Morbidelli 2003; Gomes 2003; Murray-Clay \& Chiang 2005, 2006). With the advent of the notion that the early Solar 
System may have suffered a dynamical instability (Thommes et al. 1999, Tsiganis et al. 2005, Morbidelli et al. 2007), 
the focus broadened, with the more recent theories invoking a transient phase with an eccentric orbit of Neptune 
(Levison et al. 2008, Morbidelli et al. 2008, Batygin et al. 2011, Wolff~et~al.~2012, Dawson \& Murray-Clay 2012). 

The consensus emerging from these studies is that the hot classical, resonant, scattered and detached populations (see 
Gladman et al. 2008 for the definition of these groups), formed in a massive planetesimal disk at $\lesssim$30 AU, and 
were dynamically scattered to their current orbits by migrating (and possibly eccentric) Neptune. The wide inclination 
distribution of the implanted populations indicates that Neptune's migration was long range and relatively slow, 
such that there was enough time for various processes to excite the orbital inclinations (Nesvorn\'y 2015).

The Cold Classicals (hereafter CCs), on the other hand, have low orbital inclinations ($i<5^\circ$) and several 
physical properties (ultra red colors, large binary fraction, steep size distribution of large objects,
relatively high albedos) that distinguish them from all other KBO populations. The most straightforward 
interpretation of these properties is that the CCs formed and/or dynamically evolved by different processes than other 
trans-Neptunian populations. Here we consider the possibility that the CCs formed at $>$40~AU and survived Neptune's 
early `wild days' relatively unharmed (Batygin et al. 2011, Wolff et al. 2012). 

\begin{figure}
\epsscale{0.5}
\plotone{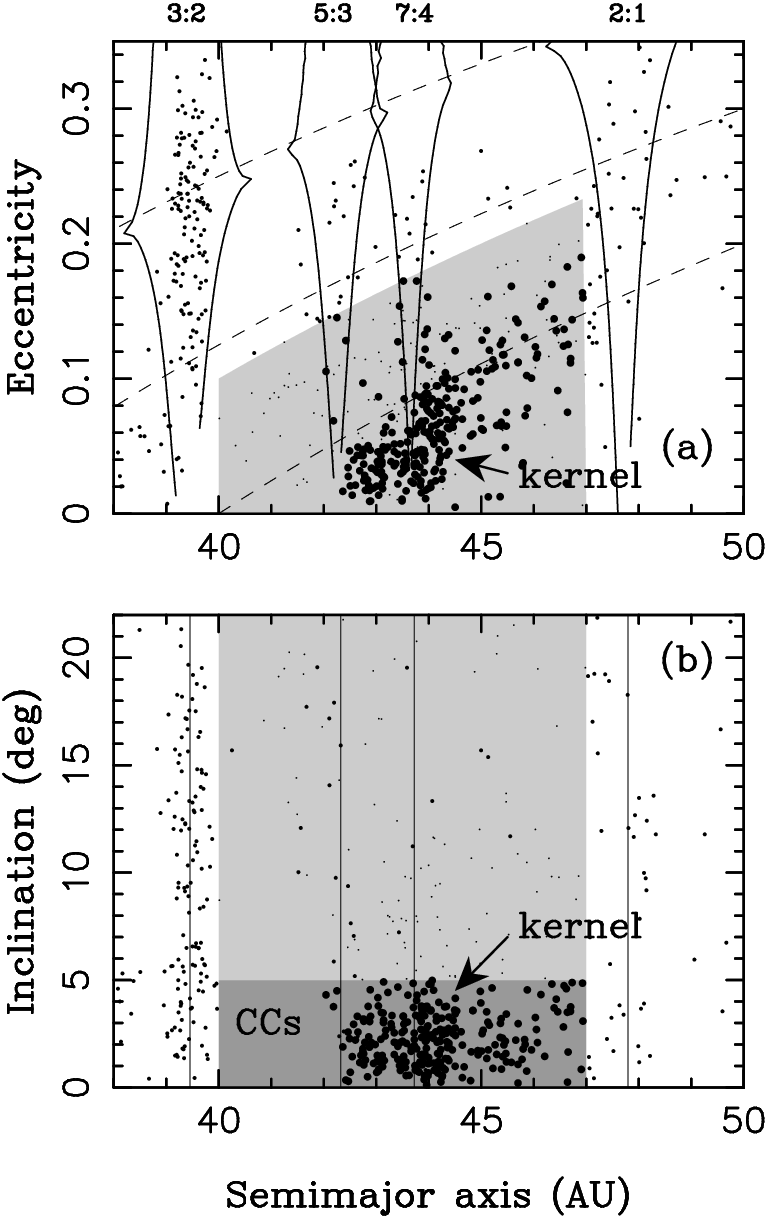}
\caption{The orbital elements of the KBOs observed in three or more oppositions. Various dynamical classes 
are highlighted. The CCs with $i<5^\circ$  are denoted by larger symbols. The solid lines 
in panel (a) follow the borders of important mean motion resonances. The low-inclination orbits
with $40<a<42$ AU are unstable due to the secular resonance overlap ($\nu_7$ and $\nu_{8}$; 
Kn\v{e}\v{z}evi\'c 1991, Duncan et al. 1995). The location of the Kuiper belt kernel is indicated
by arrows.}
\label{fig1}
\end{figure}

\begin{figure}
\epsscale{0.5}
\plotone{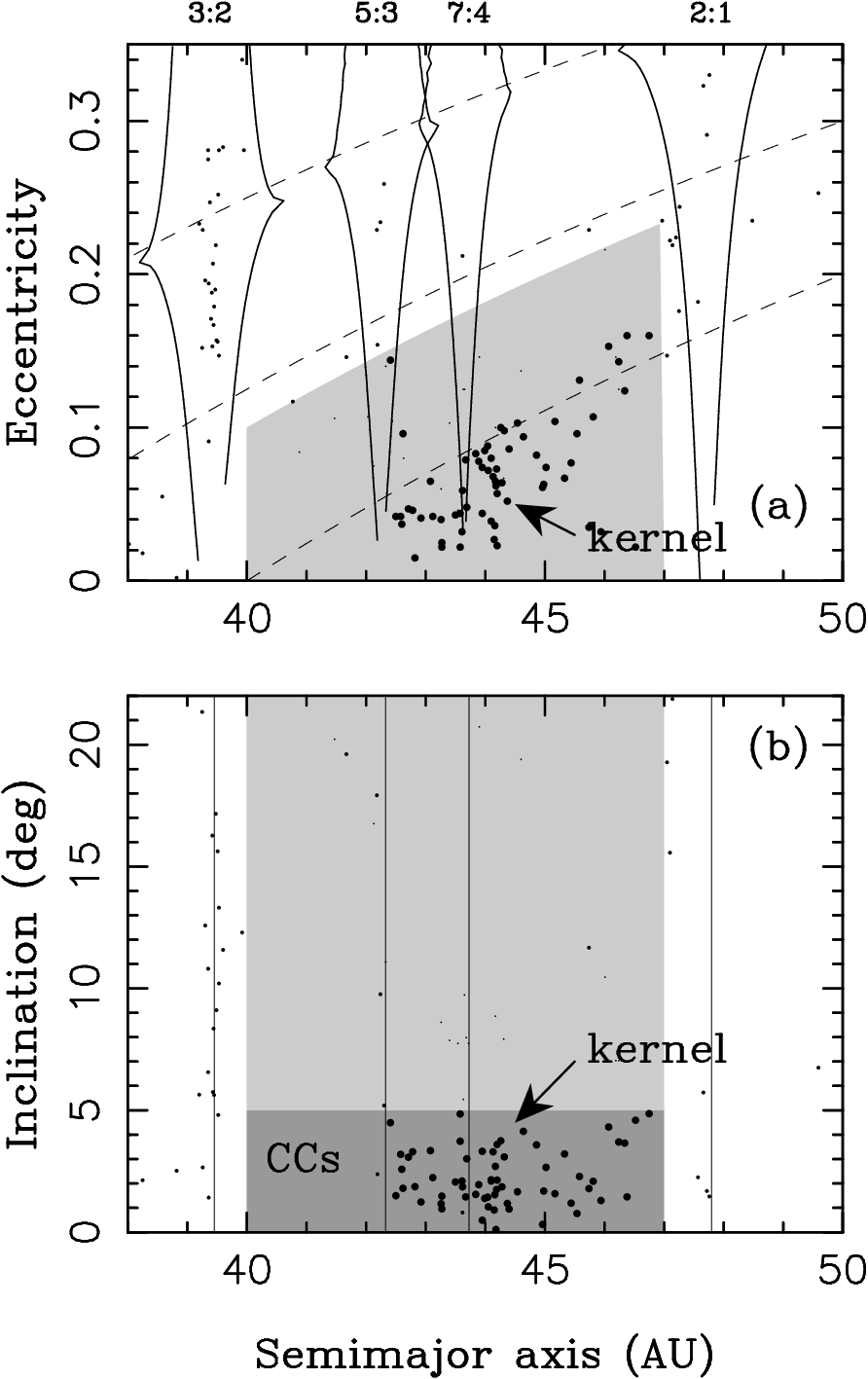}
\caption{The orbital elements of the KBOs detected by the CFEPS. The CFEPS is one of the largest surveys with 
published characterization (currently 169 objects; Petit et al. 2011). Various dynamical classes 
are highlighted. The CCs with $i<5^\circ$  are denoted by larger symbols. The solid lines 
in panel (a) follow the borders of important mean motion resonances. The location of the Kuiper belt 
kernel is indicated by arrows.}
\label{cfeps}
\end{figure}

According to Petit et al. (2011), the CC population can be divided into the `stirred' and `kernel' 
components. The stirred orbits have the semimajor axes $42<a<47$ AU, inclinations $i<5^\circ$, and 
small eccentricities with an upper limit that raises from $e\simeq0.05$ for $a=42$ AU to $e\simeq0.15$ for $a=47$ AU.
The kernel is a narrow concentration of low-inclination orbits with $a\simeq44$ AU, $e\sim0.05$, and 
a $\simeq$0.5-1 AU width in the semimajor axis. Figures \ref{fig1} and \ref{cfeps} illustrate the observed distribution 
of orbits. According to the Canada-France Ecliptic Plane Survey (CFEPS; Kavelaars et al. 2008), the number of KBOs 
in the kernel with absolute magnitude $H<8$ is $900\pm200$, which is roughly one third of the number of stirred 
objects with $H<8$ (Petit et al. 2011). The kernel is therefore an important component of the Kuiper belt (see also 
Jewitt et al. 1996). 

Chiang (2002) and Chiang et al. (2003) discussed the possibility that the concentration of bodies near 44 AU
is a collisional family (see, e.g., Nesvorn\'y et al. 2015 for a recent review of collisional families
in the asteroid belt). One problem with this hypothesis is that the collisional families in the Kuiper belt 
are expected to be stretched over many AUs in the semimajor axis (due to a relatively large contribution
of the ejection speed to the orbital motion; Marcus et al. 2011). Also, the relatively large number of $H<8$ 
objects in the kernel implies that the parent body of the putative family would have to be a dwarf 
planet at least as large as Pluto. The family hypothesis therefore seems to be improbable.

\begin{figure}
\epsscale{0.6}
\plotone{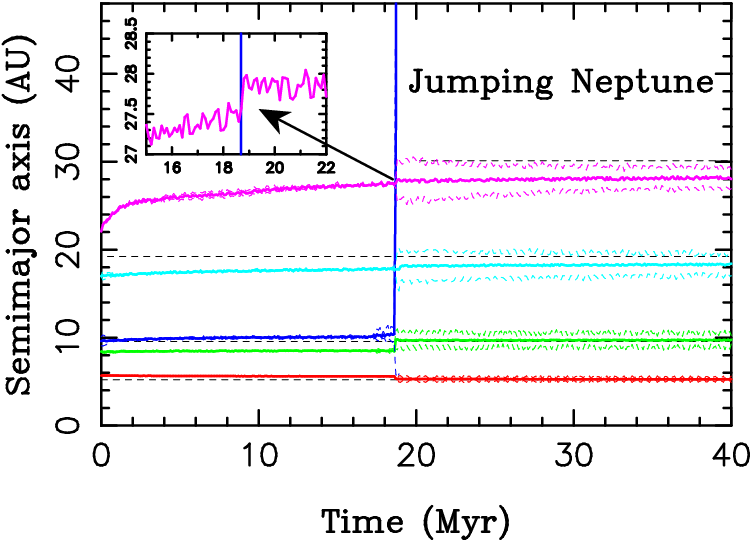}
\caption{The orbit histories of the giant planets in an instability simulation from NM12. In this example, the fifth giant 
planet was initially placed on an orbit between Saturn and Uranus and was given a mass equal to the Neptune mass. Ten 
thousand particles, representing the outer planetesimal disk, were distributed with the semimajor axis $23.5<a<29$ AU, 
surface density $\Sigma=1/a$, and low eccentricity and low inclination. With the total disk mass $M_{\rm disk}=15$ $M_{\rm Earth}$, 
each disk particle has $\simeq$0.75 Pluto mass. The plot shows the semimajor axes (solid lines), and perihelion and aphelion 
distances (thin dashed lines) of each planet's orbit in a time frame $\pm20$ Myr around the instability. Neptune migrates into 
the outer disk during the first stage of the simulation. It reaches $\simeq$27.5 AU when the instability happens ($t\simeq18.3$ Myr). 
During the instability, Neptune has a close encounter with the fifth planet and its semimajor axis jumps by 
$\simeq$0.4 AU outward (see the inset). The fifth planet is subsequently ejected from the solar system by Jupiter. Neptune's 
migration after the instability can be approximated with the e-folding timescale $\tau_2=50$ Myr.
The final orbits of the four remaining planets are a good match to those in the present Solar System (thin dashed lines).}
\label{case9}
\end{figure}

If not a family than what is the kernel? Motivated by this question, here we consider the possibility that the
kernel is a population of bodies that was captured into, and subsequently released from Neptune's 2:1 resonance. 
Our recent simulations of the planetary migration/instability in Nesvorn\'y \& Morbidelli (2012, hereafter
NM12) provide the right framework for this model. The NM12 simulations are characterized
by the first, faster migration stage during which Neptune reached $\simeq28$~AU, followed by a dynamical instability, 
when planetary encounters happened, followed by the second, very slow migration stage during which Neptune reached
$30$~AU (Fig. \ref{case9}).

We find that bodies originally at $\sim$40 AU can be captured into the 2:1 resonance during the first 
migration stage before the instability. The 2:1 resonance is at $\simeq$44 AU if the instability happens when 
Neptune is at $\simeq28$ AU. We have looked into various cases from NM12 and found that the semimajor axis of 
Neptune discontinuously changes during the instability, due to the effect of planetary encounters, by 
$\simeq$0.2-0.5 AU (Fig. \ref{case9}). Consequently, the 2:1 resonance is expected to jump by $\simeq$0.3-0.75 AU, 
which exceeds the full width of the 2:1 resonance for $e<0.1$. It is therefore expected in this model that 
resonant bodies with $e<0.1$ are released at $\simeq$44 AU, where they would remain to the present time.
\section{The Integration Method}
Our numerical integrations consist in tracking the orbits of four planets (Jupiter to Neptune) and a large number 
of test particles representing the outer planetesimal disk. To set up an integration, Jupiter, Saturn and Uranus 
were placed on their current orbits.\footnote{The dependence of the results on the orbital behavior of Jupiter, 
Saturn and Uranus was found to be minor. We determined this by comparing our nominal results with fixed orbits 
to those obtained when these planets were forced to radially migrate.} 
Neptune was placed on an orbit with the semimajor axis $a_{\rm N,0}$, eccentricity 
$e_{\rm N,0}$, and inclination $i_{\rm N,0}$. These elements define the Neptune orbit before the main stage of
migration/instability. In most of our simulations we used $a_{\rm N,0}=24$~AU, because the wide inclination 
distribution of the hot, resonant and detached population requires that Neptune's migration was long range
($a_{\rm N,0}\lesssim25$ AU; Nesvorn\'y 2015), $e_{\rm N,0}=0$ and $i_{\rm N,0}=0^\circ$.  The {\tt swift\_rmvs4} 
code (Levison \& Duncan 1994) was used to follow the evolution of planets and disk particles. The 
code was modified to include fictitious forces that mimic the radial migration and damping. These forces were 
parametrized by the exponential e-folding timescales, $\tau_a$, $\tau_e$ and $\tau_i$, where $\tau_a$ 
controls the radial migration rate, and $\tau_e$ and $\tau_i$ control the damping rate of $e$ and $i$. Here we 
set $\tau_a \sim \tau_e \sim \tau_i$ $(=\tau_1)$, because such roughly comparable timescales were suggested by 
previous work.\footnote{The precession frequencies of planets are not affected by the torques from the outer disk 
in our simulations, while in reality they were. This is an important approximation, because the orbital precession 
of Neptune can influence the degree of secular excitation of the CCs (Batygin et al. 2011).} 

The numerical integrations of the first migration stage were stopped when Neptune reached $a_{\rm N,1}=27.8$ AU. Then,
to approximate the effect of planetary encounters during the instability, we applied a discontinuous change 
of Neptune's semimajor axis and eccentricity, $\Delta a_{\rm N}$ and $\Delta e_{\rm N}$. In reality, we find
from NM12 that Neptune may suffer up to several close encounters with another planet, and that the evolution of 
$a_{\rm N}$ and $e_{\rm N}$ may be complex, with two (or more) encounters significant contributing to the
overall change of $a_{\rm N}$ and $e_{\rm N}$. Studying such evolution histories with multiple changes of  
$a_{\rm N}$ and $e_{\rm N}$ is left for future work. Here we focus on a simple case where the overall change
is approximated by a single event. Motivated  by the NM12 results, we set $\Delta a_{\rm N}=0$, 0.25, 0.5 or 0.75 AU,
and $\Delta e_{\rm N}=0$, 0.05, 0.1, or 0.15. The purpose of $\Delta a_{\rm N}=\Delta e_{\rm N}=0$ is to have a reference 
case for the comparison purposes. No change is applied to the orbital inclination of Neptune, because an 
inclination change should not critically affect the processes studied here. [The excitation of Neptune's orbital 
inclination seen in the NM12 simulations is typically not large enough to explain why the CCs have  
an inclination distribution with the characteristic width of $\simeq 2^\circ$ (Brown 2001).]    

The second migration stage starts with Neptune having the semimajor axis $a_{\rm N,2}=a_{\rm N,1}+\Delta a_{\rm N}$. 
We apply the {\tt swift\_rmvs4} code, and migrate the semimajor axis (and damp the eccentricity) on an 
e-folding timescale $\tau_2$. By fine tuning the migration amplitude the final 
semimajor axis of Neptune was adjusted to be within 0.05~AU of its current mean $a_{\rm N}=30.11$~AU, and the 
orbital period ratio, $P_{\rm N}/P_{\rm U}$, where $P_{\rm N}$ and $P_{\rm U}$ are the orbital periods of Neptune 
and Uranus, was adjusted to end up within 0.5\% of its current value ($P_{\rm N}/P_{\rm U}=1.96$). A strict 
control over the final orbits of planets is important, because it guarantees that the mean motion and secular 
resonances in the Kuiper belt region have the correct locations. All simulations 
were run to 1 Gyr. The interesting cases were extended to 4 Gyr with the standard {\tt swift\_rmvs4} code (i.e., 
without migration/damping in the 1-4 Gyr interval).

As for the specific values of $\tau_1$ and $\tau_2$ used in our integrations, we found from NM12 that the
orbital behavior of Neptune can be approximated by $\tau_1\simeq10$ Myr and $\tau_2\simeq30$~Myr for a disk mass 
$M_{\rm disk}=20$ $M_{\rm Earth}$, and $\tau_1\simeq20$ Myr and $\tau_2\simeq50$ Myr for a disk mass
$M_{\rm disk}=15$~$M_{\rm Earth}$ 
(these masses refer to the massive disk at $<30$ AU; below we consider a low-mass continuation
of this disk beyond 30 AU). Slower migration rates are possible for lower disk masses. Moreover, we find that the real
migration is not precisely exponential with the effective $\tau$ being longer than the values quoted above during the 
very late migration stages  ($\tau\gtrsim100$ Myr). This is consistent with constraints from Saturn's obliquity,
which was presumably exited by late, near-adiabatic capture in a spin-orbit resonance (e.g., Vokrouhlick\'y \& 
Nesvorn\'y 2015). Much shorter migration timescales than those quoted above do not probably apply, because
they would violate constraints from the wide inclination distribution of the hot classical and resonant 
populations (Nesvorn\'y 2015). Here we therefore use $\tau_1=10$~Myr or 30 Myr, $\tau_2=30$ or 100 Myr, and also
test a case with $\tau_2=200$ Myr.

Each simulation included 5,000 disk particles distributed from 30 to 50 AU. Their radial profile was set such that the disk 
surface density $\Sigma \propto 1/r$, where $r$ is the heliocentric distance. There is therefore an equal number
of particles (250) in each radial AU. A larger resolution is not needed, because a significant fraction of particles in the 
CC region (42-47~AU) survive, and the final statistics is therefore good enough to perform a careful comparison with observations.
The disk particles were assumed to be massless such that their gravity does not interfere with the migration/dumping 
routines. The fate of the massive disk at $<30$ AU was not studied here, because the implantation of bodies from $<30$ AU into 
the Kuiper belt was investigated elsewhere (e.g., Levison et al. 2008, Dawson \& Murray-Clay 2012, Nesvorn\'y 2015).  

An additional uncertainty relates to the dynamical structure of the original planetesimal disk at 30-50 AU. 
Here we operate under the assumption that the disk was dynamically cold with the low orbital eccentricities and low 
orbital inclinations. Since Neptune's inclination remains small in our simulations, we do not identify any dynamical effects 
that could significantly influence the orbital inclinations in the 42-47 AU region (passing mean motion 
resonances do not affect inclinations much). The initial inclinations distribution was therefore chosen to be 
similar to that inferred for the present population of the CCs from observations. Specifically, we used 
$N(i)\,{\rm d}i = \sin i \exp(-i^2/2\sigma_i^2)\,{\rm d}i$,  with $\sigma_i=2^\circ$ (Brown 2001, Gulbis et al. 2010). 
The interaction of bodies with migrating mean motion resonances is known to depend on their orbital eccentricity,
with low eccentricities more likely resulting in capture. The choice of the orbital eccentricity distribution 
could therefore, in principle, affect the results. The initial eccentricities in our simulations were distributed 
according to the Rayleigh distribution with $\sigma_e=0.01$, 0.05 or 0.1, where $\sigma_e$ is the usual 
scale parameter of the Rayleigh distribution (the mean of the Rayleigh distribution is equal to 
$\sigma_e\sqrt{\pi/2}$). Studying cases with larger eccentricity values would not be useful, because most main 
belt orbits with $e>0.1$ are dynamically unstable.
\section{Results}
We first discuss a reference simulation to illustrate the suggested relationship between the 2:1 resonance population 
deposited at $\simeq$44 AU during Neptune's jump, and the Kuiper belt kernel. Then, in Section 3.2, we  
explain how the results differ from the reference case when various model parameters are varied. The orbital
distribution of the CC orbits beyond 45 AU is discussed in Section 3.3. Finally, in Section 3.4, we deal with  
the orbital dynamics of bodies below 40 AU, including the population captured in the 3:2 resonance. We show
that the 3:2 population captured from the initial orbits with $a>30$ AU should represent only a small fraction 
($\sim1$\%) of all Plutinos.
\subsection{A Reference Case}
\begin{figure}
\epsscale{0.5}
\plotone{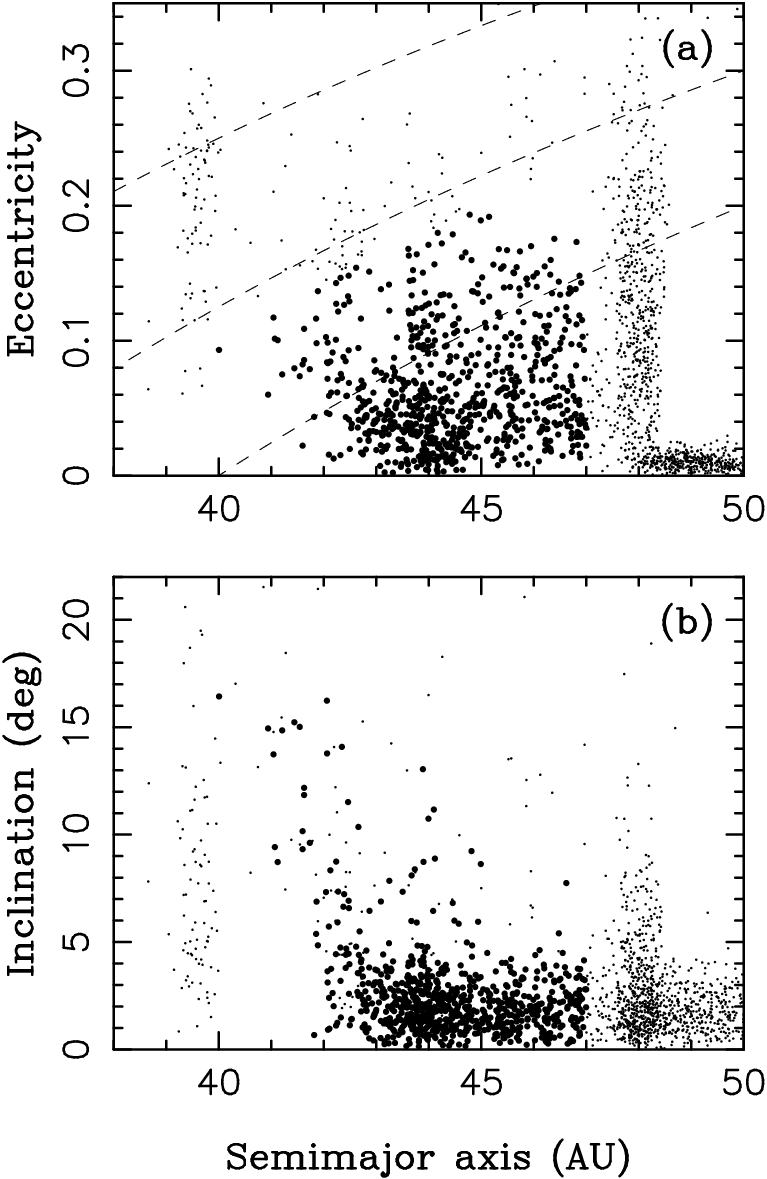}
\caption{The final distribution of orbits obtained in the simulation with $a_{\rm N,0}=24$ AU, $\tau_1=30$ Myr, 
$a_{\rm N,1}=27.8$ AU, $\Delta a_{\rm N} = 0.5$ AU, $\Delta e_{\rm N} = 0.05$, and $\tau_2=100$ Myr.
At the beginning of the simulation, 5000 test particles were distributed on low-inclination ($\sigma_i=2^\circ$) 
low-eccentricity ($\sigma_e=0.01$) orbits between 30 and 50 AU. The bold symbols denote the orbits that
ended with $40<a<47$ AU and $q=a(1-e)>36$ AU.}
\label{case1}
\end{figure}

\begin{figure}
\epsscale{0.5}
\plotone{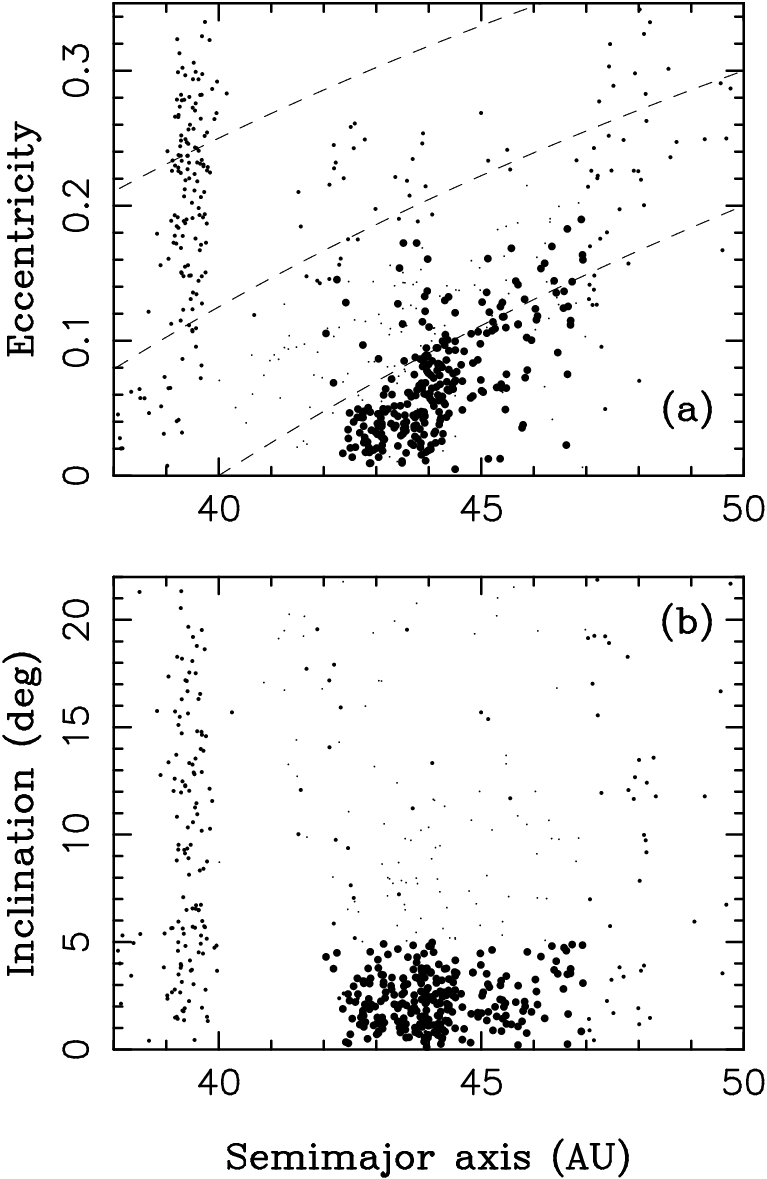}
\caption{The same as Figure \ref{fig1} but without labeling of different populations. This figure is useful 
when comparing the simulation results with observations.}
\label{realcc}
\end{figure}

Figure \ref{case1} shows the final distribution of orbits in the reference case with  $a_{\rm N,0}=24$ AU, $\tau_1=30$~Myr, 
$a_{\rm N,1}=27.8$ AU, $\Delta a_{\rm N} = 0.5$ AU, $\Delta e_{\rm N} = 0.05$, $\tau_2=100$ Myr, and $\sigma_e=0.01$. This
distribution can be compared to Figure \ref{realcc}, but we caution that Figure \ref{realcc} contains 
observational biases while Figure \ref{case1} does not. Also, in Figure \ref{case1} we highlight with larger 
symbols all final orbits with $42<a<47$ AU and $q=a(1-e)>36$ AU, while to highlight an orbit in Figure~\ref{realcc}
we also require that $i<5^\circ$ (to filter out the hot classicals). 

\begin{figure}
\epsscale{0.5}
\plotone{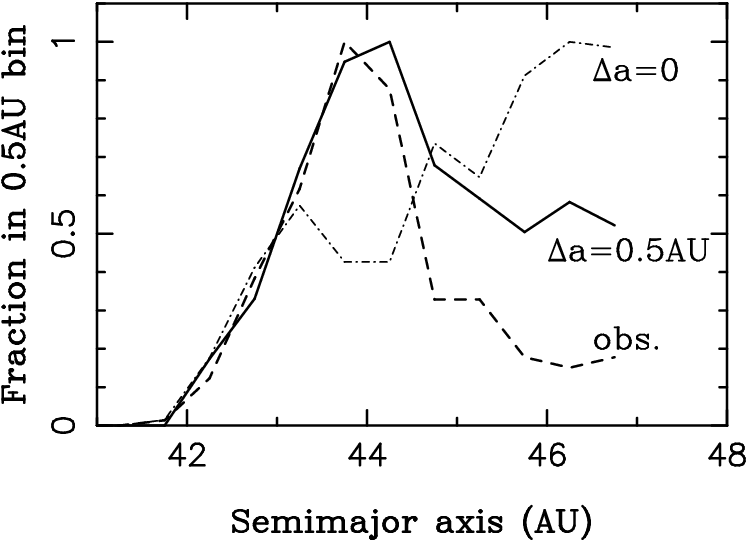}
\caption{The semimajor axis distribution of the model (solid and dot-dashed lines) and observed (dashed) KBOs. 
The observed orbits were taken from Figure 1. The model orbits were obtained in a simulation with $a_{\rm N,0}=24$ AU, 
$\tau_1=30$ Myr, $a_{\rm N,1}=27.8$~AU, $\Delta e_{\rm N} = 0.05$, $\tau_2=100$ Myr, and $\sigma_e=0.01$. 
Here we highlight the difference between cases with $\Delta a_{\rm N} = 0.5$ AU (solid line) and $\Delta a_{\rm N} 
= 0$ (dot-dashed line). The case with $\Delta a_{\rm N} = 0$ shows the orbit density increasing with the semimajor 
axis. It does not fit observations well. The case with $\Delta a = 0.5$ AU, on the other hand, shows 
a concentration of orbits at the location of the Kuiper belt kernel at 44 AU. These orbits were left behind 
by the 2:1 resonance when Neptune jumped. In both cases, we only consider orbits with $i<5^\circ$ and 
$q=a(1-e)>36$ AU.} 
\label{histo1}
\end{figure}

The distributions of orbits in Figures \ref{realcc} and \ref{case1} share several similarities, but also 
show several important differences. First, the model distribution of the CC orbits in Figure \ref{case1} 
is clustered at $a\simeq44$ AU and $e<0.1$. This is the orbital location of the kernel. In the model, the 
concentration of orbits was created by the 2:1 resonance, which collected captured bodies before
Neptune's jump, and then released them with $a\simeq44$ AU and $e<0.1$, when Neptune jumped. This can be 
conveniently demonstrated by comparing these model results with a simulation where $\Delta a_{\rm N} = 0$. 
Figure \ref{histo1} shows a comparison of various semimajor axis distributions. The observed distribution 
raises from below 42 AU, where very few low-inclination object exist due to the presence of overlapping secular
resonances, toward the location of the kernel at $\simeq$44 AU. The model distribution obtained with $\Delta 
a_{\rm N} = 0.5$ AU reproduces this trend very well, while that with $\Delta a_{\rm N} = 0$ AU does not. This
happens essentially because no concentration of orbits is created at $\simeq$44 AU if Neptune migrates 
smoothly, that is without a jump. 

\begin{figure}
\epsscale{0.8}
\plotone{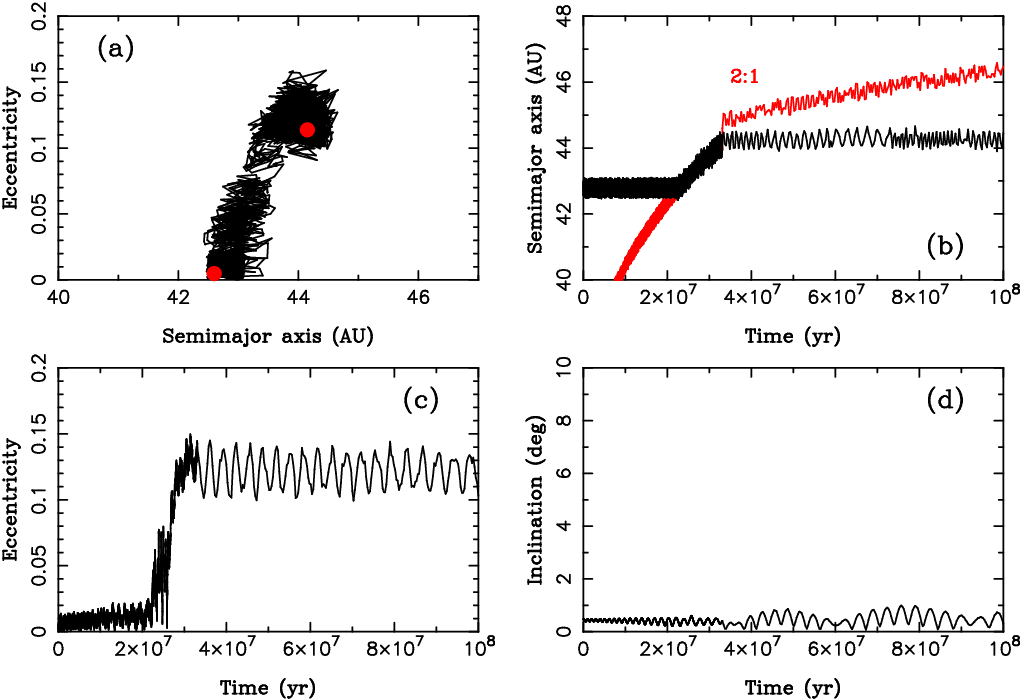}
\caption{The orbital history of a test particle that was released from the 2:1 resonance when Neptune jumped
(at $t=32.5$ Myr) 
The panels show: the (a) path of the disk particle in the $(a,e)$ projection; the two red dots show the initial 
and final orbits, (b) semimajor axis, (c) eccentricity, and (d) inclination. From $t\simeq22$ Myr to $t\simeq32$ 
Myr, the 2:1 resonance angle, $\sigma_{2:1}=2\lambda-\lambda_{\rm N}-\varpi_{\rm N}$, librates with a full amplitude 
of $\simeq200^\circ$.}
\label{tp1}
\end{figure}

Figure \ref{tp1} shows an example of orbit taken from our simulation with $\Delta a_{\rm N} = 0.5$ AU. This example 
illustrates how an objects starting below 43 AU is captured by the the 2:1 resonance, which transports it to $\simeq$44 AU, 
where the body is released from the resonance during Neptune's jump. This is a typical evolution path followed by 
bodies deposited into the kernel region in our simulations.

\begin{figure}
\epsscale{0.8}
\plotone{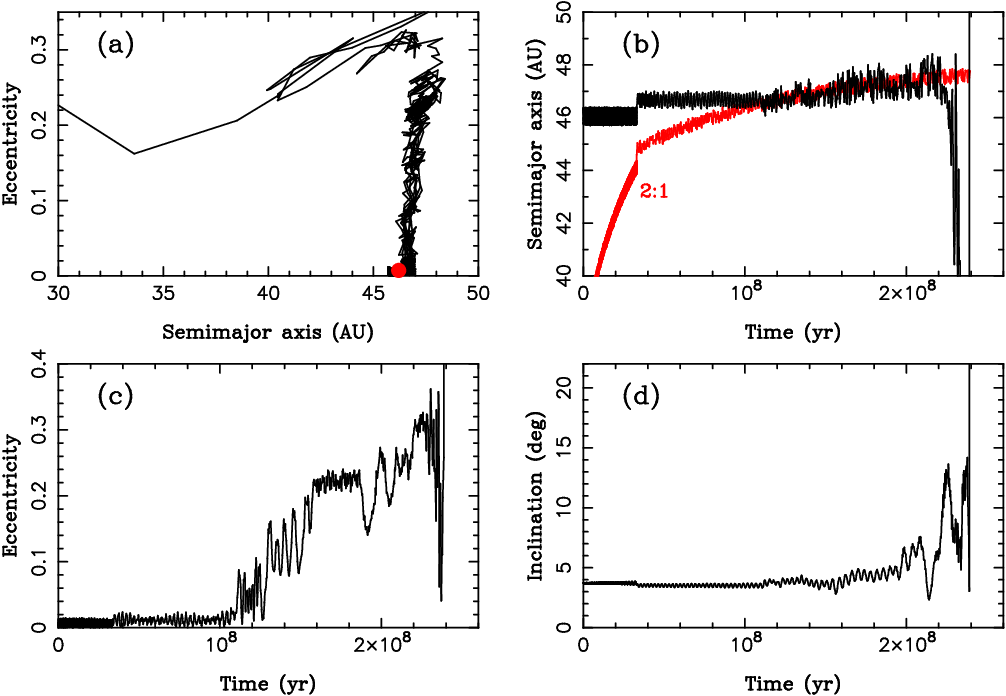}
\caption{The orbital history of a test particle that started with $a\simeq46$ AU and was destabilized by the  
2:1 resonance. The panels show: the (a) path of the disk particle in the $(a,e)$ projection; the red dot shows 
the initial orbit, (b) semimajor axis, (c) eccentricity, and (d) inclination. The particle was captured 
into the 2:1 resonance at $t\simeq110$~Myr after the start of the simulation. It remained on a resonant orbit with
a very large libration amplitude to $t\simeq220$ Myr, and subsequently evolved on a Neptune-crossing orbit.}
\label{tp2}
\end{figure}

The distribution of orbits for $a\gtrsim 45$ AU presents a challenge. The semimajor axis distribution of the
observed orbits in Figures \ref{realcc} and \ref{histo1} sharply drops from 44 AU to 45 AU. The density
of known low-$i$ orbits between 45 and 47 AU is roughly 15-30\% lower than the peak value at 44 AU. 
This is often described as the Kuiper belt ``edge'' or ``cliff'' (Bernstein et al. 2004). The edge is not well 
reproduced in our simulations. The model density obtained with $\tau_2=100$ Myr 
drops only to $\simeq$50-60\% of the peak value. In the model, the depletion is caused by a slowly migrating 
2:1 resonance that captures and removes bodies from this region (see Figure \ref{tp2} for an illustration). 
To obtain a better agreement, one would thus either need to create a stronger concentration of orbits at 
$\simeq$44 AU, such that the contrast increases by a factor of two or so, or produce a more severe depletion 
in the 45-47 AU region. 

While this could be potentially achieved by slowing down the migration (see Section 3.2), a more fundamental problem 
with the result in Figure \ref{case1} is that the 2:1 resonance captures objects from the 45-47~AU region and becomes 
overpopulated, relative to observations, at the end of the simulation (see the clump of low-$i$ resonant orbits at 
$\simeq 48$ AU in Figure \ref{case1}). This happens because only some of the orbits captured in the 2:1 resonance 
become destabilized later, as in Figure \ref{tp2}. Most captured orbits are stable and survive inside the 
resonance. Also, the low-$i$ and low-$e$ orbits beyond the 2:1 resonance ($a \gtrsim 49$ AU) remain practically 
unchanged in our simulations, while no bodies on such orbits were detected so far. We discuss this problem in 
Section 3.3.
\subsection{Dependence on Model Parameters}
\begin{figure}
\epsscale{0.4}
\plotone{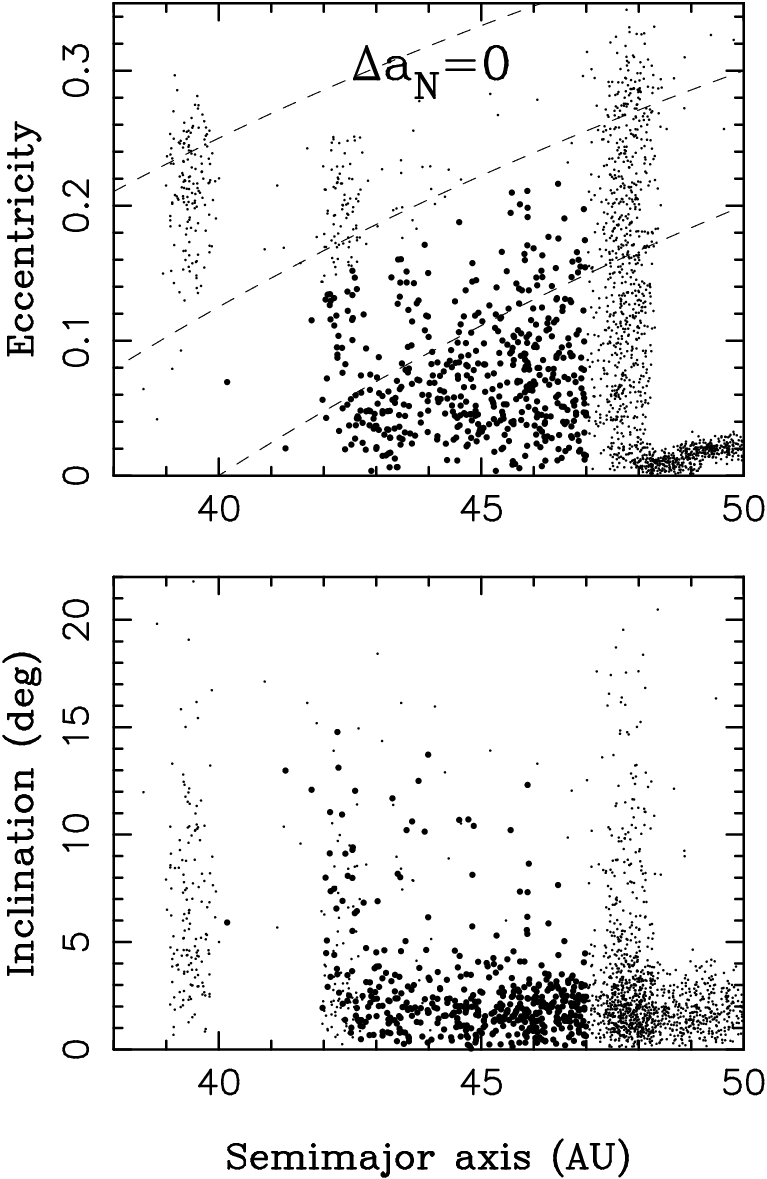}
\plotone{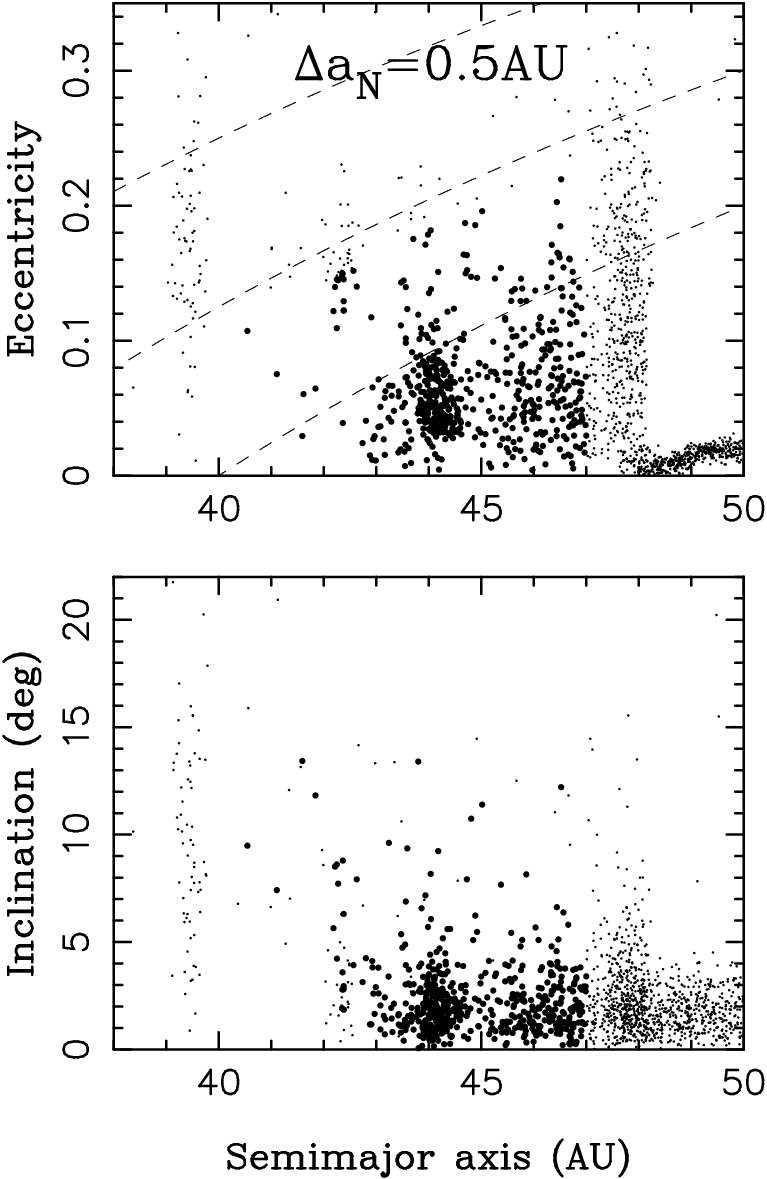}
\caption{The final distribution of orbits obtained in the simulation with $a_{\rm N,0}=24$ AU, $\tau_1=30$ Myr, 
$a_{\rm N,1}=27.8$ AU, $\Delta e_{\rm N} = 0.1$, and $\tau_2=100$ Myr. The panels on the left show the result 
for $\Delta a_{\rm N} = 0$, while those on the right show the result for $\Delta a_{\rm N} = 0.5$ AU.
The concentration of orbits at $\simeq$44 AU in the right panels was created by the 2:1 resonance when
Neptune jumped. At the beginning of the simulation, 5000 test particles were distributed on low-inclination 
($\sigma_i=2^\circ$) low-eccentricity ($\sigma_e=0.01$) orbits between 30 and 50 AU. The bold symbols denote the 
orbits that ended with $40<a<47$ AU and $q=a(1-e)>36$ AU.}
\label{case1ab}
\end{figure}

\begin{figure}
\epsscale{0.4}
\plotone{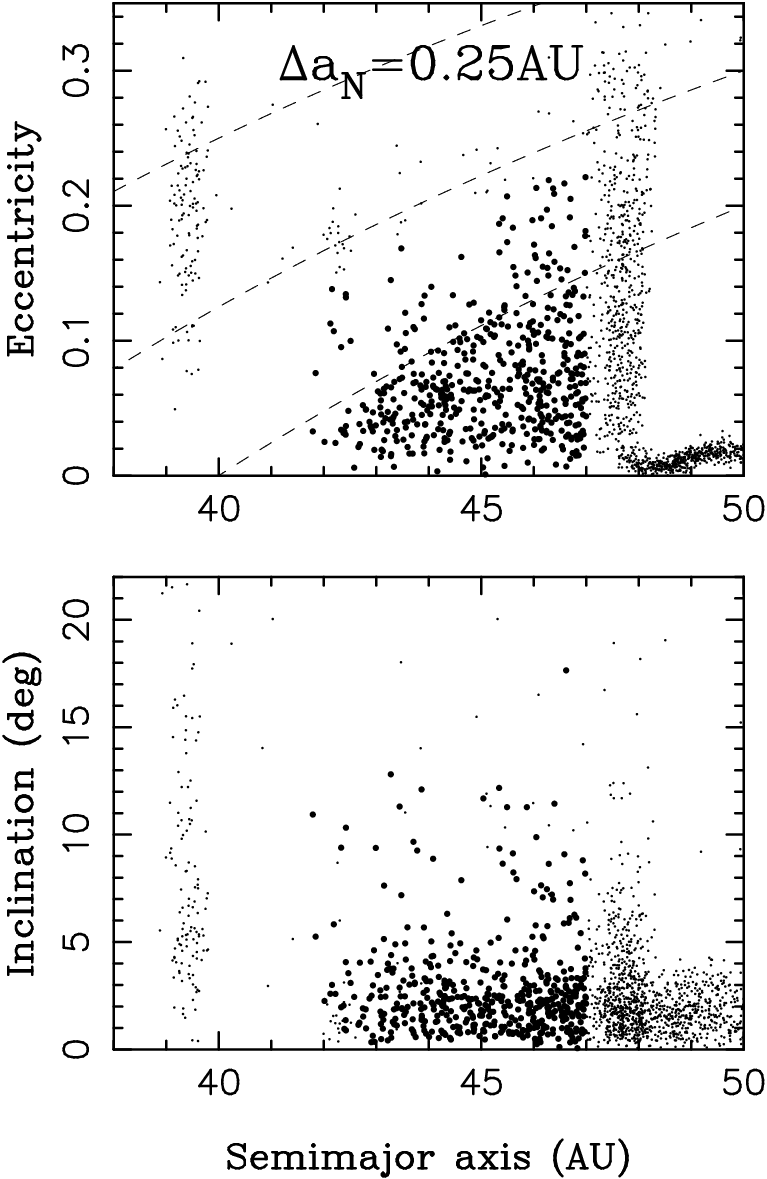}
\plotone{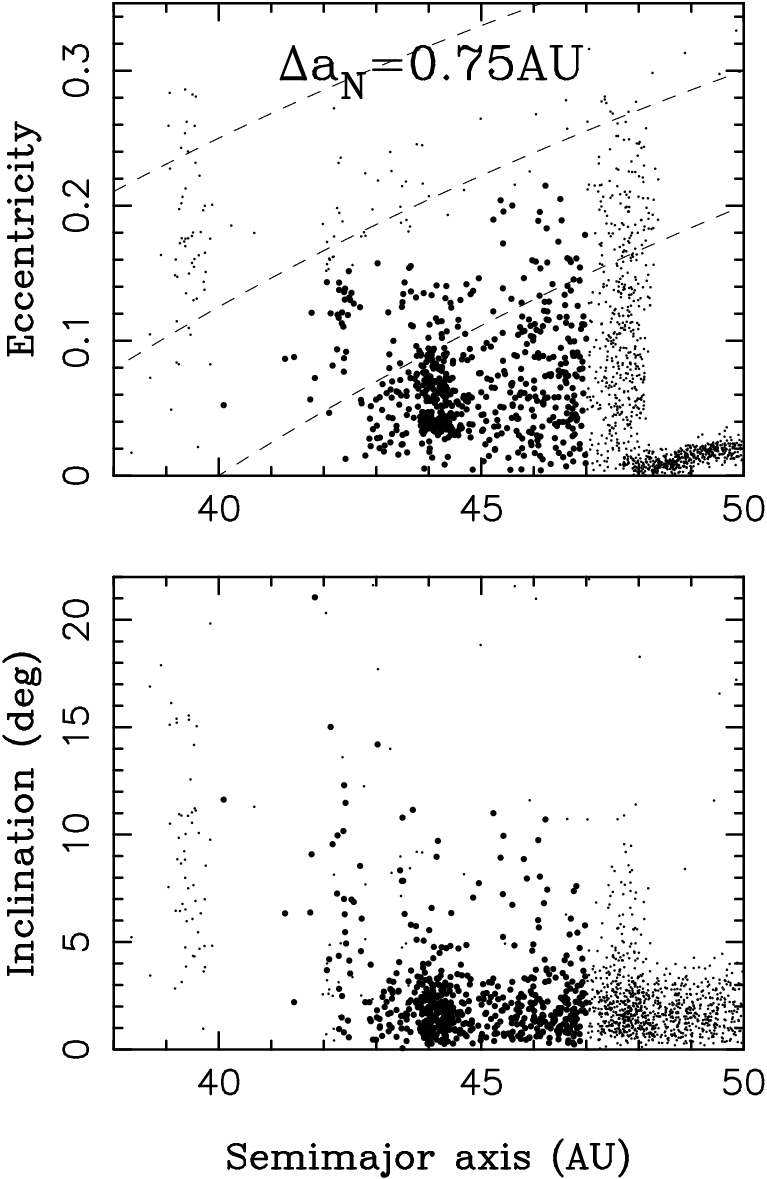}
\caption{The final distribution of orbits obtained in the simulation with $a_{\rm N,0}=24$ AU, $\tau_1=30$ Myr, 
$a_{\rm N,1}=27.8$ AU, $\Delta e_{\rm N} = 0.1$, and $\tau_2=100$ Myr. The panels on the left show the result 
for $\Delta a_{\rm N} = 0.25$ AU, while those on the right show the result for $\Delta a_{\rm N} = 0.75$ AU.
The concentration of orbits at $\simeq$44 AU in the right panels was created by the 2:1 resonance when
Neptune jumped. The bold symbols denote the orbits that ended with $40<a<47$ AU and $q=a(1-e)>36$ AU.}
\label{dan}
\end{figure}

\begin{figure}
\epsscale{0.4}
\plotone{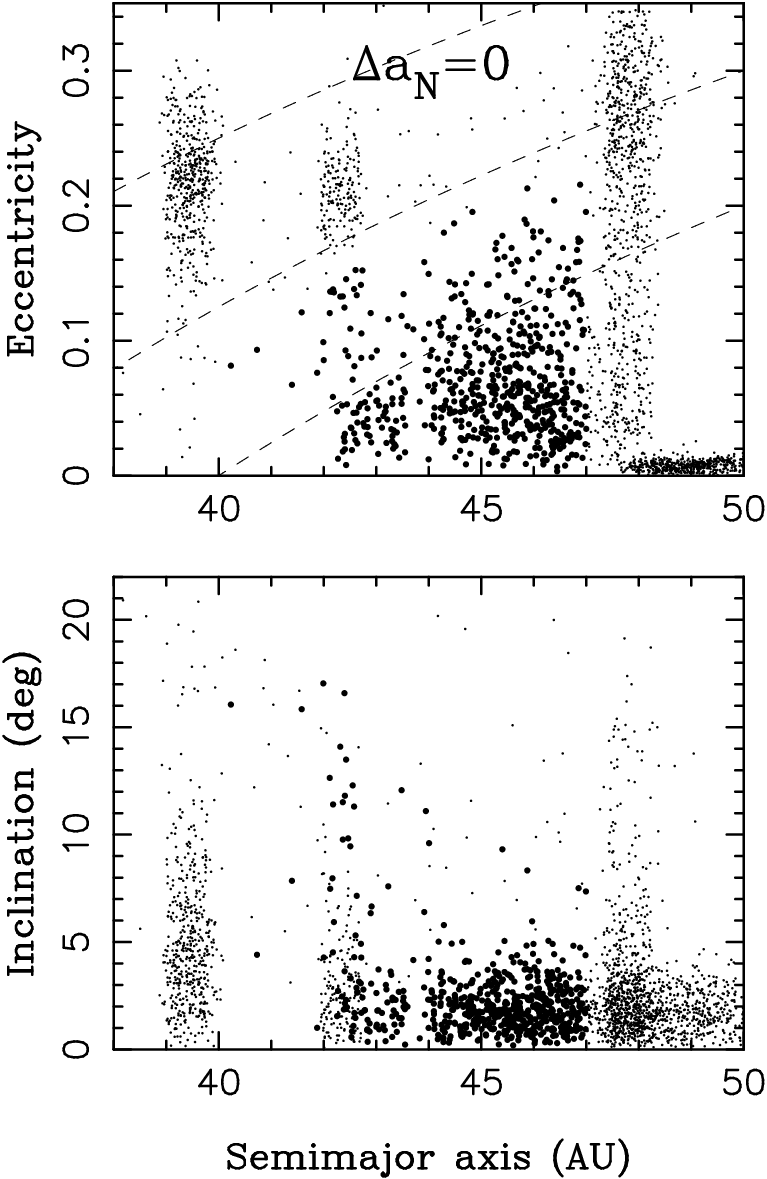}
\plotone{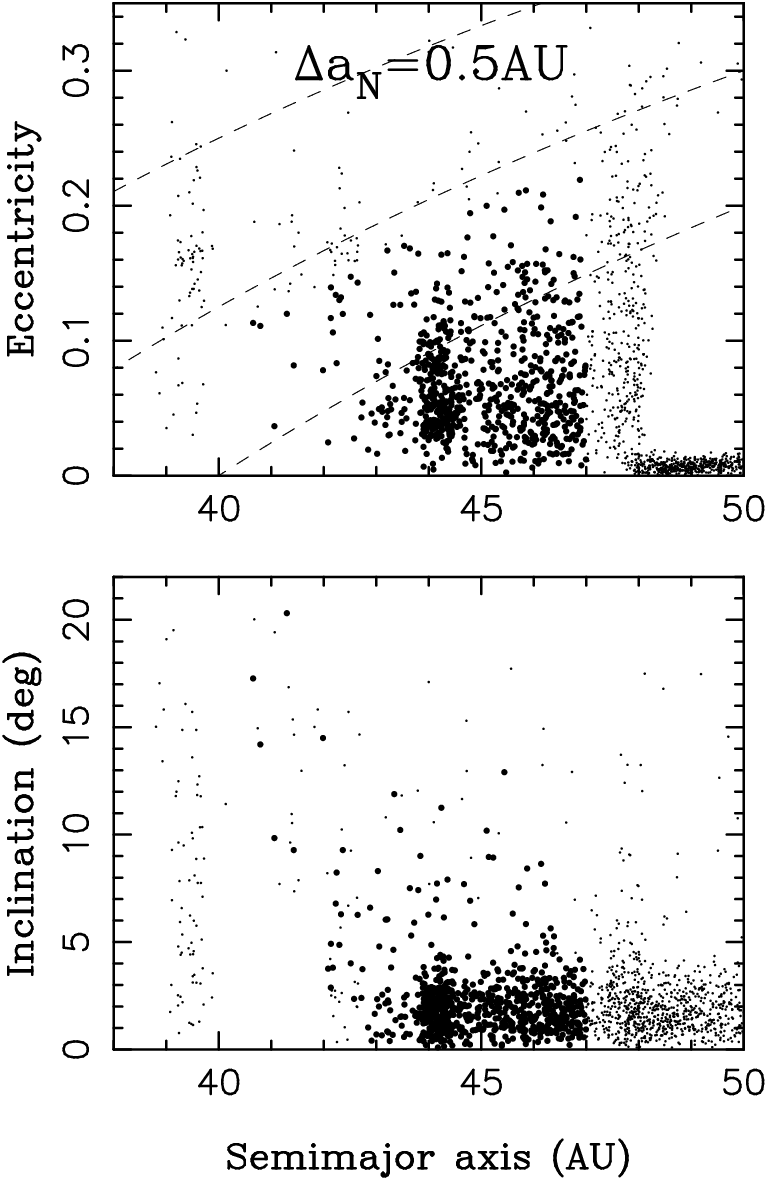}
\caption{The final distribution of orbits obtained in the simulation with $a_{\rm N,0}=24$ AU, $\tau_1=10$ Myr, 
$a_{\rm N,1}=27.8$ AU, $\Delta e_{\rm N} = 0.1$, and $\tau_2=30$ Myr. The panels on the left show the result 
for $\Delta a_{\rm N} = 0$, while those on the right show the result for $\Delta a_{\rm N} = 0.5$ AU.
The bold symbols denote the orbits that ended with $40<a<47$ AU and $q=a(1-e)>36$ AU.}
\label{case2}
\end{figure}

\begin{figure}
\epsscale{0.4}
\plotone{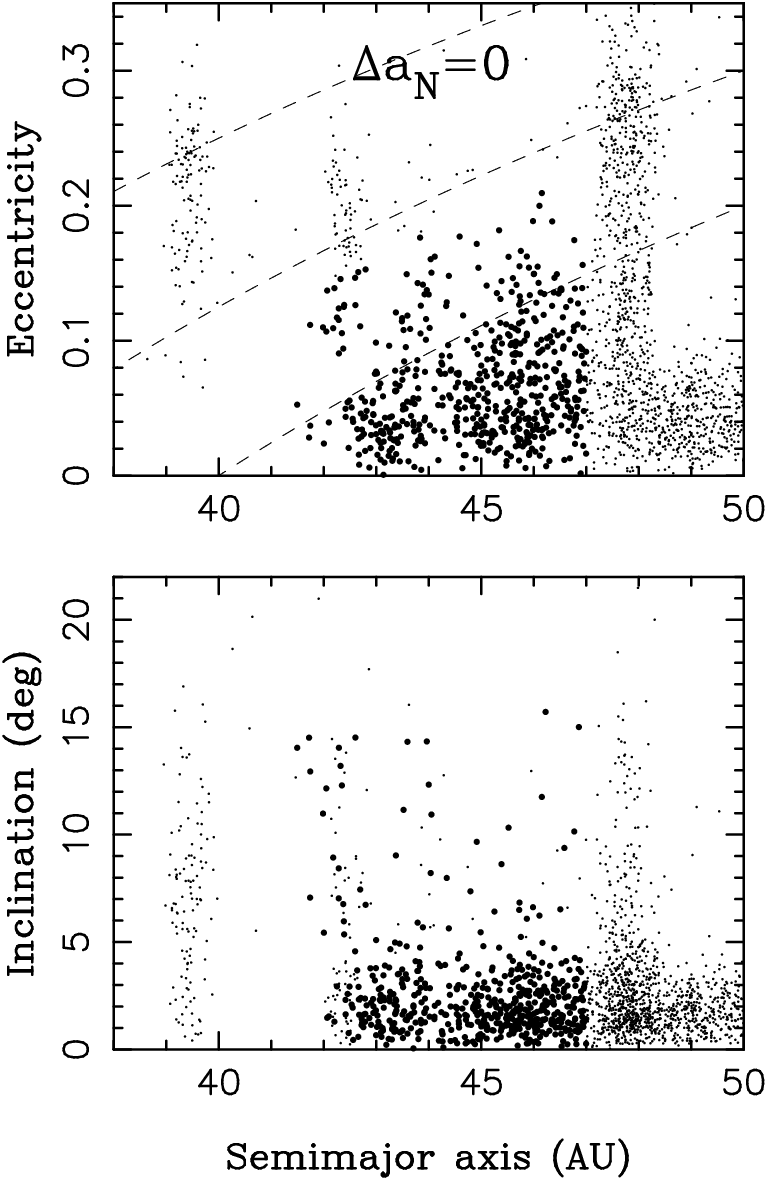}
\plotone{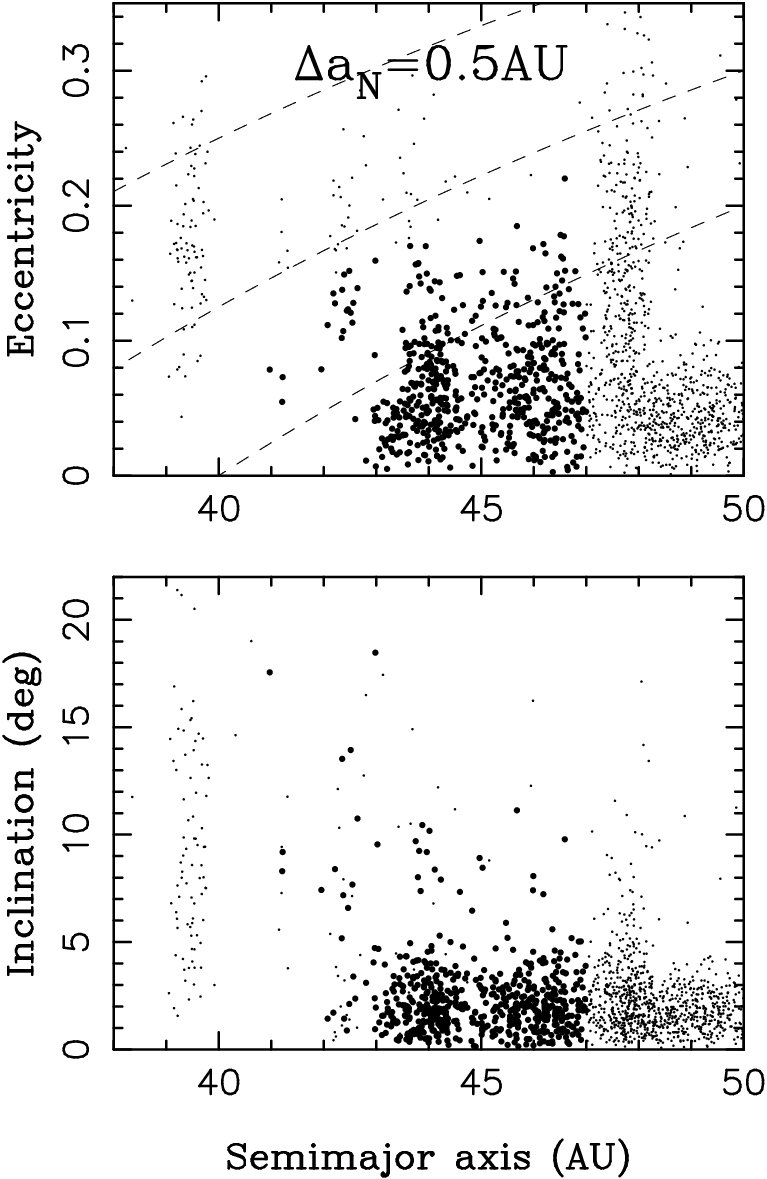}
\caption{The same as Figure \ref{case1ab} but for the test particles having larger initial eccentricities 
($\sigma_e=0.05$).}
\label{case1cd}
\end{figure}

Here we discuss the dependence of the results on: (1) $\Delta e_{\rm N}$ (Figure \ref{case1ab}), (2) $\Delta a_{\rm N}$
(Figure \ref{dan}), (3) $\tau_1$ and $\tau_2$ (Figure~\ref{case2}), and (4) the initial eccentricities of particles 
in the 30-50 AU region (Figure~\ref{case1cd}).

As for (1), we used $\Delta e_{\rm N}=0$, 0.05, 0.1 and 0.15. The results for $\Delta e_{\rm N}=0.05$ were discussed
in the previous section. Figure \ref{case1ab} illustrates the result for the same model parameters as in the 
previous section, but $\Delta e_{\rm N}=0.1$. In this case, the kernel population obtained with $\Delta a = 0.5$ AU 
(right panels in Fig. \ref{case1ab}) is somewhat more concentrated near 44 AU than it was in Figure \ref{case1}. 
The origin of this difference is not understood. It may have something to do with with the dynamics of large libration 
amplitude orbits inside the 2:1 resonance, and its dependence on $e_{\rm N}$. With $\Delta e_{\rm N}=0.1$ the kernel 
orbits obtained in the model have the semimajor axis between 43.8 and 44.6 AU, and eccentricity between 0.03 and 0.09,
in a very close match to the distribution of the kernel orbits inferred from the CFEPS observations (Petit et al. 2011; 
their Figure 4). The left panels in Fig. \ref{case1ab} show that the kernel does not form if Neptune does not jump, 
as expected. 

The results obtained with $\Delta e_{\rm N}=0$ (and $\Delta a_{\rm N}=0.5$ AU) also show the formation of the kernel,
but in this case, if $\sigma_e=0.01$, a low-$e$ segment of the original disk survives at 44-45 AU. These orbits 
are not altered by the 2:1 resonance, which just jumps over them. Since such orbits are not observed, either 
$\Delta e_{\rm N}>0$ or $\sigma_e>0.01$. The case with $\Delta e_{\rm N}=0.15$ also does not apply, because the 
CC population at 42-45 AU is disrupted when Neptune's eccentricity becomes large (here we used $\tau_2=100$ Myr; 
shorter migration timescales could work better with $\Delta e_{\rm N}=0.15$, but see Wolff et al. (2012),
Dawson \& Murray-Clay (2012), and Morbidelli et al. (2014)). This tests indicate that the preferred value for 
the eccentricity change of Neptune's orbit is $\Delta e_{\rm N}\simeq0.05$-0.1. 

As for (2), Figure \ref{dan} shows the result for $\Delta a_{\rm N}=0.25$ AU and 0.75 AU ($\Delta e_{\rm N}=0.1$
and $\tau_2=100$ Myr). The case with $\Delta a_{\rm N}=0.25$ AU is within the range of the NM12 results, 
while in the one with $\Delta a_{\rm N}=0.75$~AU, Neptune's jump is probably too large to be realistic 
(the cases with $\Delta a_{\rm N}>0.5$ AU do not happen too often in NM12). We used $\Delta a_{\rm N}=0.75$ AU
just in case some future modification of the NM12 simulation setup would lead to a larger jump of Neptune.
The results obtained here for $\Delta a_{\rm N}=0.25$ AU do not lead to the formation of the kernel (Figure \ref{dan}, 
left panels), and are in fact very similar to those obtained with $\Delta a_{\rm N}=0$. The ones obtained with 
$\Delta a_{\rm N}=0.75$ AU, on the other hand, are very similar to the previous case with $\Delta a_{\rm N}=0.5$ 
AU, where a well-defined kernel forms. We conclude that the kernel constraint requires that Neptune's semimajor 
axis changed by $\simeq$0.5-0.75 AU (with the lower values in this range being more in line with the NM12 results).

Figure \ref{case2} illustrates the dependence of the results on the migration timescale. Here we assumed that
$\tau_1=10$ Myr and $\tau_2=30$ Myr, and left all other model parameters from Figure~\ref{case1ab} unchanged.   
These shorter migration timescales would be more appropriate if Neptune's migration was driven by a more massive
planetesimal disk. The results are similar to those obtained for longer migration timescales.
The kernel forms for $\Delta a = 0.5$~AU and does not form for $\Delta a = 0$. The compact orbital structure of 
the kernel obtained for $\Delta a = 0.5$ AU is very similar to that obtained in the case with the longer migration 
timescales. This shows that the results are not sensitive to the migration timescale of Neptune. A minor 
difference between Figures \ref{case2} and \ref{case1ab} can be noticed in the 45-47 AU region, where the
population of orbits is less depleted if the migration is faster. This is related to the migration-speed
dependence of the removal by the 2:1 resonance. With $\tau_2=200$ Myr, which was the longest migration
timescale considered here, only $\sim$32\% of bodies survived in the 45-47 AU region (while 74\% survived for 
$\tau_2=30$ Myr, and 47\% survived for $\tau_2=100$ Myr). 

Finally, we simulated several cases, where the test particles at 30-50 AU were given a wider initial eccentricity 
distribution. We considered cases with the Rayleigh distribution of eccentricities and $\sigma_e=0.05$ or 0.1. 
Figure \ref{case1cd} shows the result obtained with $\sigma_e=0.05$ (also $\tau_1=30$ Myr,
$\tau_2=100$ Myr, and $\Delta e_{\rm N}=0.1$). Relative to Figures \ref{case1ab} and \ref{case2} the 
concentration of orbits near 44 AU obtained with $\Delta a_{\rm N}=0.5$ AU is more fuzzy. The results 
obtained for $\sigma_e=0.1$ are similar. This was expected, because the interaction of orbits with the 2:1 resonance 
is more stochastic if the orbits have significant eccentricities. We conclude the kernel population 
is expected to be more fuzzy if the disk particles are assumed to have larger orbital eccentricities.

It is not clear which of these results fit the existing data best. This is in part due to the fact that 
the existing observations do not characterize the kernel population very well. Also, we are not motivated to 
attempt any detailed fits yet, because the NM12 instability simulations show that Neptune's semimajor axis may 
have suffered more than one important jump (because there was more than one important planetary encounter). If 
that was the case, the exact orbital structure of the kernel would depend on the exact sequence of jumps
and their magnitude. A detailed investigation into these issues goes beyond the scope of the model presented 
here, where we approximated the dynamical instability by a single event. Our goal in this work is to show 
that the kernel can plausibly be explained if Neptune jumped during the instability, and that this explanation 
does not depend on fine tuning of the model parameters (except that Neptune's jump had to occur near 28 AU, 
such that the kernel was deposited by the 2:1 resonance near 44 AU). 
\subsection{The Kuiper Belt Edge}
The low density of the low-$i$ orbits at 45-47 AU and in the 2:1 resonance, and the lack of low-$i$ orbit 
object detection beyond $\simeq$49 AU are probably part of the same issue. As we discussed in Sections 3.1 
and 3.2, the 2:1 resonance can help to deplete the region at 45-47 AU, especially if Neptune's migration
was slow, but this does not resolve the problem, because the low-$i$ orbits accumulate in the 2:1 resonance 
and those beyond the current location of the 2:1 resonance remain essentially intact.   
We investigated two potential solutions to this problem. First, we considered the possibility that the Kuiper 
belt observations are incomplete and the orbital region in question is in fact populated by bodies that so far 
avoided detection. Second, we looked into the expected orbital distribution while assuming that the original 
planetesimal disk had a sharp edge, perhaps due to the inefficiency in the accretion of bodies beyond 
$\sim$45 AU. We consider the reference case from Section 3.1 for these tests.

We used the CFEPS simulator to understand the detection statistics. The simulator was developed by the CFEPS team to aid 
the interpretation of their observations (Kavelaars et al. 2009). Given intrinsic orbital and magnitude distributions, 
the CFEPS simulator returns a sample of objects that would be detected by the survey, accounting for the flux biases, 
pointing history, rate cuts and object leakage (Kavelaars et al. 2009). In the present work, we input our 
model populations in the simulator to compute the detection statistics. We then compare the orbital distribution 
of the detected objects with the actual CFEPS detections shown in Figure \ref{cfeps}.

The CFEPS simulator takes as an input the orbital element distribution from our numerical model and absolute 
magnitude ($H$) distribution. The magnitude distribution was taken from Fraser et al. (2014). It was assumed 
to be described by a broken power law with 
$N(H)\,{\rm d}H=10^{\alpha_1(H-H_0)}\,{\rm d}H$ for $H<H_{\rm B}$ 
and  
$N(H)\,{\rm d}H=10^{\alpha_2(H-H_0)+(\alpha_1-\alpha_2)(H_{\rm B}-H_0)}\,{\rm d}H$ for $H>H_{\rm B}$,
where $\alpha_1$ and $\alpha_2$ are the power-law slopes for objects brighter and fainter than 
the transition, or break magnitude $H_{\rm B}$, and $H_0$ is a normalization constant.  Fraser et al. (2014) 
reported that $\alpha_1\simeq 1.5$, $\alpha_2 \sim 0.2$ and $H_{\rm B}=6.9$ represent the best fit to the
detection statistics of the CCs. We used these values as a reference, and varied them within the error uncertainty
given by Fraser et al. (2014) to understand the sensitivity of the results on various assumptions. 

The basic result from these tests, assuming Fraser's magnitude distribution, is that the problem with low
orbit density beyond $\simeq$45 AU cannot be resolved by invoking the observational incompleteness (and 
nothing else). This is because the low inclination bodies with $H\simeq6.9$ are readily detected by the CFEPS 
even for perihelion distances $q\simeq50$ AU (Petit et al. 2012). Therefore, there must have been 
a real edge of the original disk beyond which no large bodies formed, or sizable bodies formed but for 
some reason they were fewer in number, or smaller in size, than those that formed at $<$45 AU, and are
not detected by the current surveys. 

As a follow-up on the latter option, we experimented with the magnitude distributions where the overall number 
of bodies was assumed to drop toward 50 AU, or the number of bodies was kept fixed, but $H_{\rm B}$ was 
assumed to rise toward 50 AU. For simplicity, we assumed that $\alpha_1$ and $\alpha_2$ were unchanging with 
the heliocentric distance. These magnitude distribution assumptions were applied to the objects in the 
{\it original} disk, and were then propagated to the final orbital distributions.  The goal of these tests was 
to understand whether a {\it gradual} change in the number of bodies or the break magnitude in the original 
disk can produce a distribution that would be consistent with present observations, or whether a sharper 
transition is needed. 

\begin{figure}
\epsscale{0.5}
\plotone{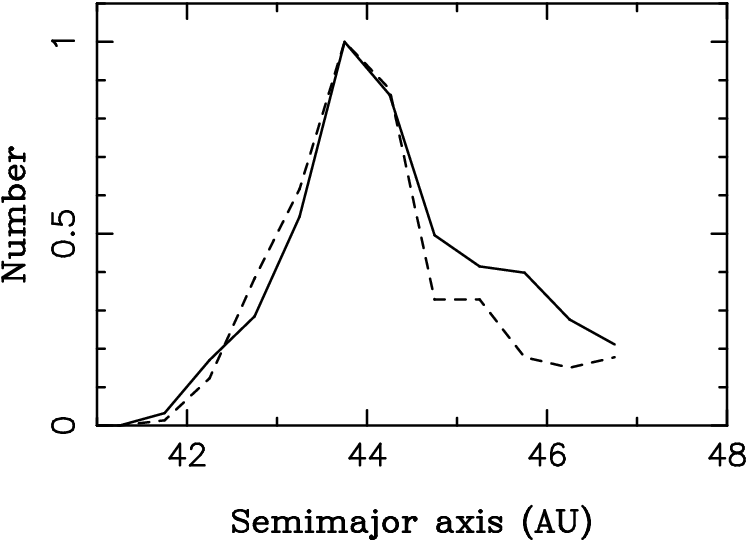}
\caption{The semimajor axis distribution of the model (solid line) and observed (dashed) bodies. The model
orbits were obtained with $a_{\rm N,0}=24$ AU, $\tau_1=30$ Myr, $a_{\rm N,1}=27.8$ AU, $\Delta a_{\rm N} = 0.5$ AU, 
$\Delta e_{\rm N} = 0.05$, and $\tau_2=100$ Myr. The CFEPS simulator was applied here. We assumed that the number
density of objects with $H<9$ per semimajor axis interval was constant up to 45 AU, and dropped linearly 
from 45 AU to zero at 50 AU.}
\label{varn}
\end{figure}

Figure \ref{varn} shows the semimajor axis distribution obtained for the same case shown in Figure \ref{histo1}
(for $\Delta a_{\rm N}=0.5$ AU),
but with variable $N_{\rm B}$, where $N_{\rm B}$ is the number of bodies with $H<H_{\rm B}=6.9$. Specifically, 
$N_{\rm B}$ was assumed to be constant for the heliocentric distance $r<45$ AU, and linearly decrease for 
$r>45$ AU such that $N_{\rm B}=0$ at 50 AU. We propagated this distribution to the final result of our simulation 
and applied the CFEPS simulator to it. The number of objects was adjusted such that the number of detections 
in the main belt was comparable to the number of actual CFEPS detections. The agreement in Figure \ref{varn}
is reasonable. The model distribution now drops toward 47 AU in very much the same way the real distribution does.
Also, the model distribution shows no detections beyond the 2:1 resonance, and only a very few objects inside 
the resonance, which also agrees quite nicely with observations.

\begin{figure}
\epsscale{0.5}
\plotone{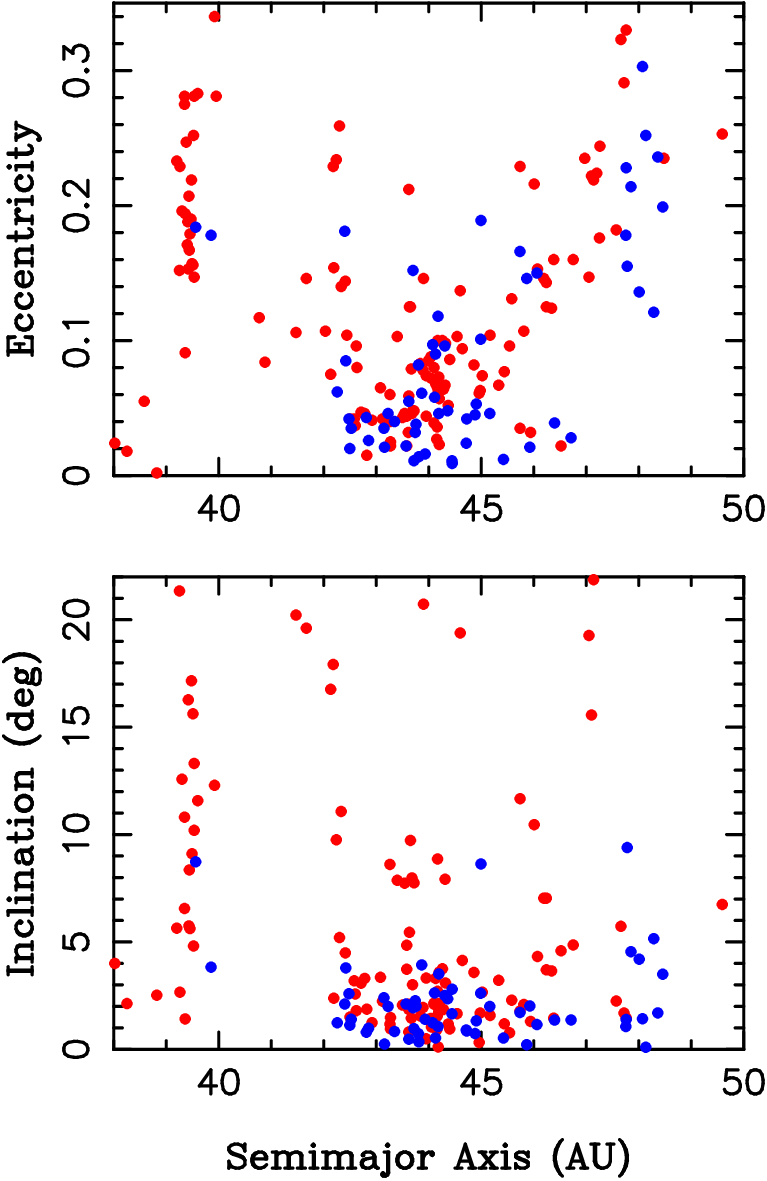}
\caption{A comparison of the observed (red symbols) and model (blue) orbit distributions. The model distribution 
was obtained in the simulation with $a_{\rm N,0}=24$ AU, $\tau_1=30$ Myr, $a_{\rm N,1}=27.8$ AU, $\Delta a_{\rm N} 
= 0.5$ AU, $\Delta e_{\rm N} = 0.05$, and $\tau_2=100$ Myr. Here we assumed that the break magnitude $H_b$ in 
the original disk was 6.9 for $r \leq 45$ AU, and dropped linearly with $r$ to $H_b=9$ at 50 AU.}
\label{varh}
\end{figure}

Figure \ref{varh} shows the orbital distribution obtained with $H_{\rm B}=6.9$ for $r<45$~AU, and 
$H_{\rm B}$ linearly increasing from $H_{\rm B}=6.9$ at $r=45$ AU to $H_{\rm B}=9$ at $r=50$ AU. The number of
objects brighter than the break magnitude, $N_{\rm B}$, was kept fixed in this case. The correspondence to
observations in Figure \ref{varh} is reasonable. All detected objects with low orbital eccentricities are located 
in the 42-47 AU region. Several bodies detected in the 2:1 resonance have $e\gtrsim0.15$ and $i\lesssim5^\circ$,
just as observed. There are fewer detected bodies in the 3:2 resonance and along the $q\simeq40$ AU line for 
$45<a<47.5$ AU relative to actual detections, but this can be explained if some of the actual detections are 
low-$i$ interlopers from the population implanted into the Kuiper belt from $r<30$ AU (Nesvorn\'y 2015).

We tested many different combinations of the prescription for $N_{\rm B}(r)$ and $H_{\rm B}(r)$, and found that
the results became unsatisfactory when the drop of $N_{\rm B}(r)$ or the rise of $H_{\rm B}(r)$ was more 
gradual than the ones discussed above. The underlying condition is dictated by the requirement that the disk 
bodies at 50 AU are either too few or too faint to be detected. Since the existing CFEPS is sensitive to 
$H\lesssim8.5$ bodies on low-$i$ orbits at 50 AU, this implies that the number of bodies with $H\lesssim8.5$ must 
be relatively low. A more rigorous statistical analysis of this problem is left for future work.

\begin{figure}
\epsscale{0.5}
\plotone{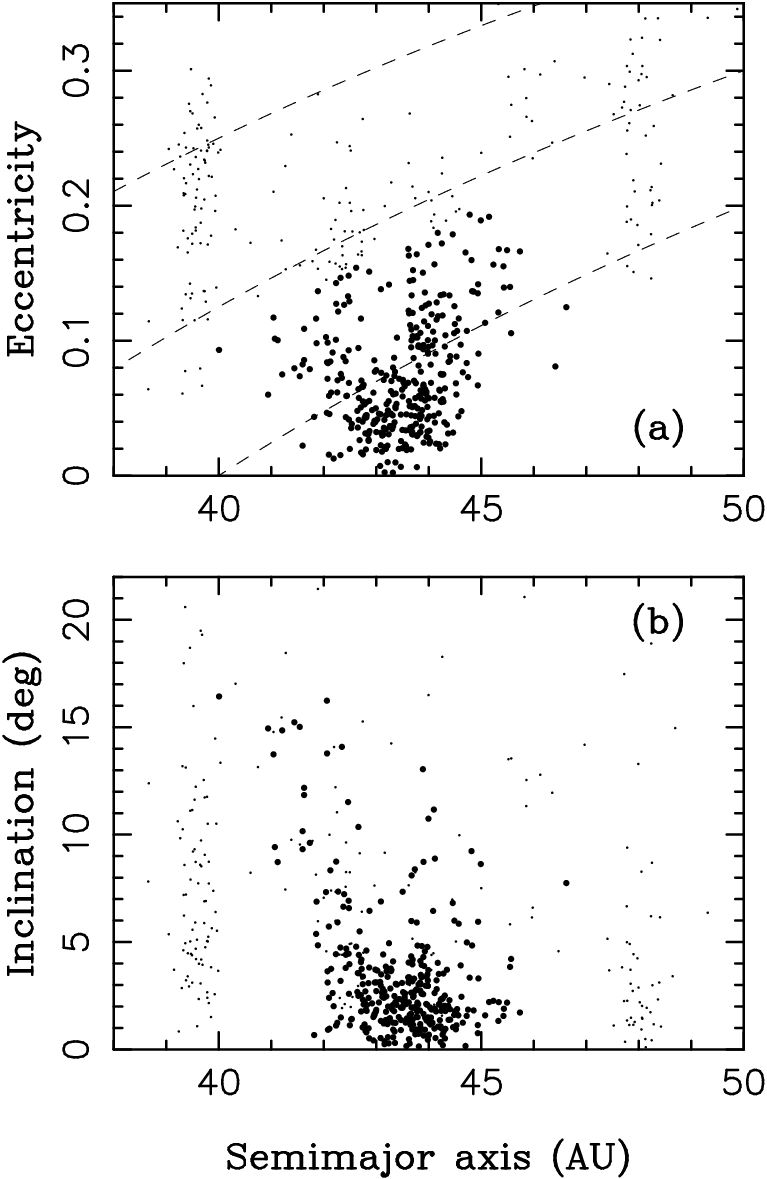}
\caption{The final distribution of orbits obtained in the simulation with $a_{\rm N,0}=24$ AU, $\tau_1=30$ Myr, 
$a_{\rm N,1}=27.8$ AU, $\Delta a_{\rm N} = 0.5$ AU, $\Delta e_{\rm N} = 0.05$, and $\tau_2=100$ Myr. Here we assumed 
that the initial disk had an outer edge at 44 AU, and there were no bodies initially located beyond 44 AU.}
\label{cut44}
\end{figure}

Another interesting possibility to consider is the original disk with a sharp edge at radius $r_{\rm edge}$ 
and no bodies beyond $r_{\rm edge}$. Figure \ref{cut44} shows the final orbital distribution of particles 
that started with $r<r_{\rm edge}=44$ AU in our simulations. The transition at $\simeq$45 AU is much sharper 
than in Figure \ref{case1} with most bodies beyond 45 AU being located in the 2:1 resonance. These bodies started 
with $r<44$ AU, were captured into the 2:1 resonance, and evolved onto resonant orbits with $e\gtrsim0.15$. 
Their orbital inclinations remained small. The stirred CC population with at 45-47 AU is not well reproduced in 
this model, but otherwise the overall distribution of bodies in Figures \ref{realcc} and \ref{cut44} is 
similar (note that no survey simulator was applied in Figure \ref{cut44}). 
\subsection{Orbital Distribution at 30-40 AU}
The original disk at 30-40 AU is completely disrupted and only a small fraction
of the original objects survive at the end of the simulations. Most of these survivors ended up in Neptune's 
3:2 resonance. Our initial concern with this result was that the observed 3:2 resonance population does not have 
a prominent low-$i$ component with physical properties that would be noticeably different from the rest. 
To state this differently, the bimodality of the inclination distribution and physical properties in the 
classical main belt does not have a counterpart among Plutinos. 

\begin{figure}
\epsscale{0.8}
\plotone{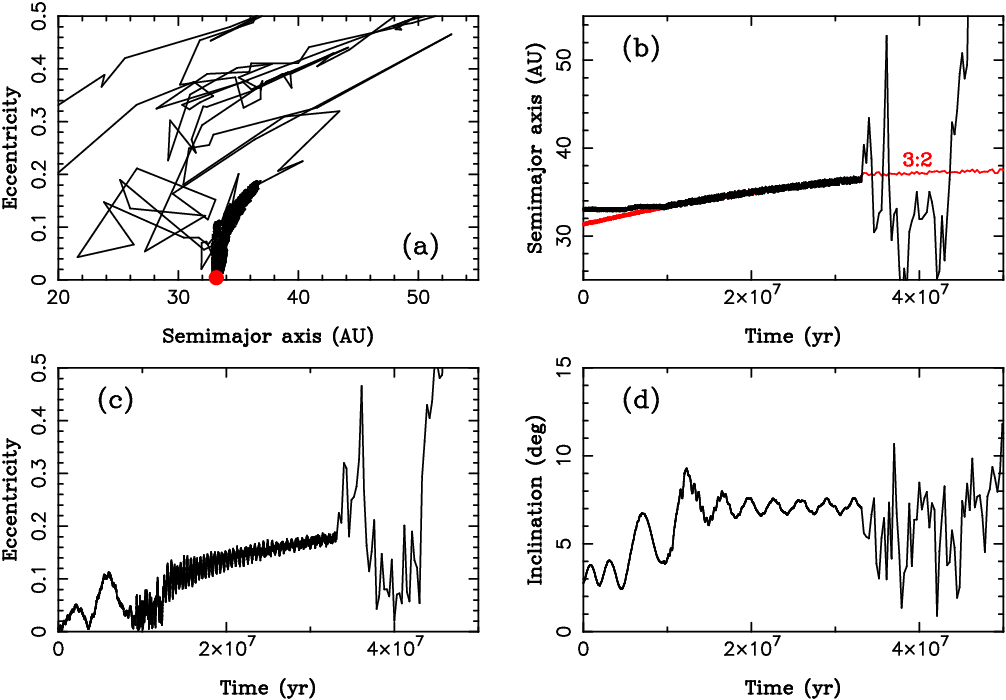}
\caption{The orbital history of a test particle that was captured into the 3:2 resonance with Neptune, and
was released from the resonance when Neptune jumped at $t=32.5$~Myr. The particle ended up on the Neptune-crossing orbit. 
The panels show: the (a) path of the disk particle in the $(a,e)$ projection; the red dot shows the initial 
orbit, (b) semimajor axis, (c) eccentricity, and (d) inclination. The 3:2 resonant angle, 
$\sigma_{3:2}=3\lambda-2\lambda_{\rm N}-\varpi_{\rm N}$, librates in the interval between $t\simeq10$ Myr 
and $t\simeq33$ Myr.}
\label{tp3}
\end{figure}

This issue goes back to the standard ideas about the effects of the long-range migration of Neptune on a dynamically 
cold disk at $>30$ AU. As Neptune migrates, the low-inclination orbits are collected in the 3:2 resonance, and 
should be found in the resonance today, while in fact they are not. We find that two factors contribute to mitigate 
this problem in our simulations. First, the population of orbits captured in the 3:2 resonance before Neptune's jump 
is released from the resonance when Neptune jumps (see Figure \ref{tp3} for an illustration). This reduces, by a 
factor of $\simeq$2-6, the number of resonant objects that remain in the 3:2 resonance at the end of our 
simulations (we find this by comparing the results of simulations with and without Neptune's jump; the reduction
factor positively correlates with the migration speed). Those that survive were captured after Neptune's jump. 

\begin{figure}
\epsscale{0.8}
\plotone{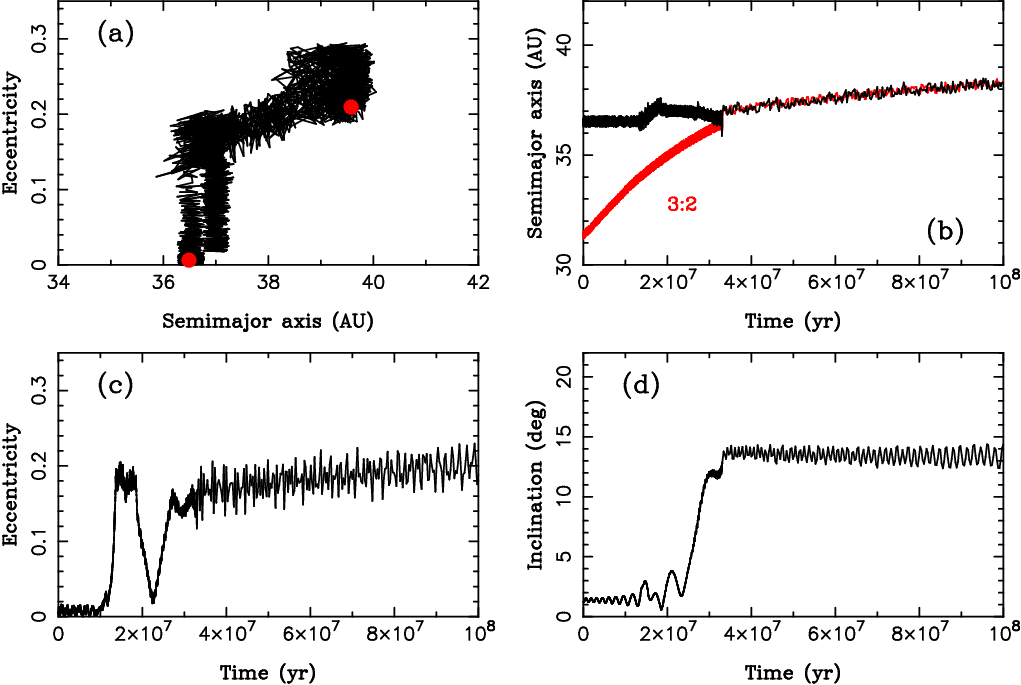}
\caption{The orbital history of a test particle that was captured into the 3:2 resonance and remained in the
resonance during the whole simulation. The orbital inclination started low and was excited to $i>10^\circ$
before the particle was captured into the 3:2 resonance. The panels show: the (a) path of the disk particle in the $(a,e)$ 
projection; the two red dots show the initial and final orbits, (b) semimajor axis, (c) eccentricity, and (d) 
inclination.}
\label{tp4}
\end{figure}

Second, Neptune's migration is very slow after the jump. As the 3:2 resonance slowly sweeps through the disk, the 
secular resonances $\nu_8$ and $\nu_{18}$, located at a slightly larger semimajor axis than the 3:2 resonance, 
accompany this motion and excite orbits to large eccentricities and large inclinations before they could enter
into the 3:2 resonance. This creates sort of a shield on outside of the migrating 3:2 resonance. Only a 
small fraction of bodies are seen to penetrate this shield and be captured in the resonance. Those that do have 
their orbital inclinations excited by the passage through $\nu_{18}$ (see Figure \ref{tp4} for an example). 
The final inclination distribution of the resonant orbits is therefore relatively wide (see, e.g., panel (b) 
in Figure \ref{case1}).\footnote{On a related note, 
there is only one equal-size large-separation binary (also red, high albedo) known in the 3:2 resonance (2007 
TY4; Sheppard et al. 2012). The heliocentric orbit of 2007 TY4 has the inclination $i\simeq11^\circ$, which is well
within the range of the inclination distribution shown in Figure \ref{case1}b. 2007 TY4 is thus probably a 
survivor of the primordial population of objects beyond 30 AU (with $r\simeq37$-39 AU being its most plausible
formation location).} 
  
In the reference case discussed in Section 3.1 (with $\Delta a_{\rm N}=0.5$ AU), we find that only $\simeq$70 particles 
ended up on stable orbits in the 3:2 resonance, while $\simeq650$ survived at $42<a<47$ AU. The fraction of bodies 
captured in the 3:2 resonance from the disk at $r>30$ AU, relative to those surviving in the main belt, is thus 
roughly 1/10. Other cases studied here show similar results. This implies that the population of bodies implanted in 
the 3:2 resonance from the disk at $r>30$~AU should be only a very small fraction of the present population of Plutinos. 
Indeed, existing observations indicate that the intrinsic population of Plutinos is $\sim$10 times larger than that 
of the CCs (e.g., Fraser et al. 2014, Adams et al. 2014). Combining the two factors of $\sim$10 discussed above, we 
therefore conclude that only one in $\sim$100 Plutinos should have physical characteristics similar to the CCs (red color, 
high albedo, binarity). This currently represents only a few objects with good orbits (the detection bias toward the 
low-$i$ orbits being factored in here). Future observations should be able to test this prediction. 
\section{Discussion}
The primordial planetesimal disk between 30 and 50 AU can be divided into two parts. The inner part of the 
disk at 30-40 AU is disrupted during Neptune's migration, with most of the surviving bodies being 
captured in the 3:2 resonance. If Neptune's migration was relatively slow ($\tau \gtrsim 10$ Myr), as required 
from the inclination distribution of the HCs (Nesvorn\'y 2015), and Neptune jumped by a fraction of AU during the 
instability, we estimate that only $\sim$1\% of today's Plutinos should have originated from the 30-40 AU region, 
while $\sim$99\% were captured from the massive disk below 30 AU. The orbital inclinations of Plutinos captured 
from 30-40 AU are expected to be substantial, because orbits are excited by the secular resonances before 
they can be captured into the 3:2 resonance. Faster migration rates and/or the absence of Neptune's jump would lead 
to a larger proportion of Plutinos being captured from the 30-40 AU region, and a narrower inclination distribution 
of the captured population. This would contradict observations, because Plutinos do not have a low-$i$ component 
with noticeably different physical properties.  

The outer disk at 40-50 AU is somewhat excited and depleted in our simulations, but not by a large factor. The 
objects originally at 40-44 AU can be captured by the migrating 2:1 resonance and released at $\simeq$44 AU when Neptune 
jumped. The orbital characteristics of this population resemble that of the kernel, a previously unexplained 
feature of the Kuiper belt. This gives support to the migration model, where Neptune slowly migrates and suffers one 
or a few scattering encounters when at $\simeq$28 AU. As a result of the encounter(s), Neptune's semimajor axis 
changes by $\simeq$0.5 AU, and then continues migrating, very slowly, toward its present value of 30.1 AU. 
Neptune's 2:1 resonance, moving past 45 AU during the late stage of migration, removes $\simeq$25-70\% of 
bodies and contributes to the orbital depletion at 45-47~AU. The effect of the 2:1 resonance cannot explain, however, 
on its own, the lack of low inclination orbits at $>47$ AU. Several possible solutions to this problem were discussed 
in Section 3.3.     

The inclination excitation in the 42-47 AU region is minimal in our simulations with most bodies retaining orbits 
with $i<5^\circ$. Only a few dozens of orbits is seen to evolve to $i>5^\circ$. This implies that the contamination
of the HCs from the cold disk at 42-47 AU should be minimal. We can roughly estimate the contamination factor 
from our simulations. The number of orbits with $i>5^\circ$ at 42-47 AU is roughly 1/20 of those surviving with 
$i<5^\circ$ at 42-47 AU. This shows that the contamination should be of the order of 5\% relative to the CC population.
Fraser et al. (2014) estimated that the CC population represents only $\sim$1/30 of the HC populations. Considering these
two factors together, only 1 in $\sim$600 main belt objects with $i>5^\circ$ would be expected to have started with
$i<5^\circ$ and $42<a<47$~AU. The real contamination factor could be larger, for example, if the disk
at 42-47 AU was dynamically heated by some early process, or was excited later by the secular interactions with 
the fifth planet (Batygin et al. 2012; NM12).

Initially, there were 1250 particles in the 42-47 AU region in our simulations. The number of particles remaining 
in this region at the end of the simulations is between 550 and 700. The dynamical depletion factor is therefore only 
$\sim$2. Fraser et al. (2014) found that the total mass of the CCs population is $\sim 3\times10^{-4}$ $M_{\rm Earth}$. 
We can therefore estimate that the original mass in the 42-47 AU region was $\sim 6\times10^{-4}$ $M_{\rm Earth}$, 
most of which was probably concentrated below $\simeq$45 AU (see the discussion in Section 3.3). This roughly implies 
$\sim 2\times10^{-4}$ $M_{\rm Earth}$ in each radial AU, or the surface density of solids $\Sigma_{\rm s} \sim 2 \times 10^{-5}$ 
g cm$^{-2}$. This is $\sim$4 orders of magnitude smaller than the surface density needed to form sizable objects in
the standard coagulation model (Kenyon et al. 2008). It is possible that the original surface density was much higher 
and bodies were removed by fragmentation during collisions (Pan \& Sari 2005), but the presence of loosely
bound binaries places a strong constraint on how much mass can be removed (Nesvorn\'y et al. 2011). 

A possible solution of this problem is that the CCs formed by a gravitational collapse of solids that were locally 
concentrated by their interaction with the gaseous nebula. For example, Youdin \& Goodman (2005) suggested that large planetesimals 
can form from the concentrations of particles produced by the streaming instability. Numerical simulations give support 
to this model and show that the streaming instability can operate in a low-mass environment assuming that the local 
metalicity can be slightly increased over the solar metalicity (Johansen et al. 2009). Since only a small fraction of 
the available mass may be converted into sizable planetesimals by this process, the original surface density in the 42-47 AU 
region could have been higher than inferred above. The large binary fraction among the CCs ($>$30\%; Noll et al. 2007, 
2014) can provide an evidence for the gravitational collapse model, because binaries are expected to form if the 
collapsing clouds have important rotation (Nesvorn\'y et al. 2010).

\acknowledgments
This work was supported by NASA's Outer Planet Research (OPR) program.
All CPU-expensive simulations in this work were performed on NASA's Pleiades 
Supercomputer.\footnote{http://www.nas.nasa.gov/hecc/resources/pleiades.html} 
We thank Alessandro Morbidelli and David Vokrouhlick\'y for helpful discussions. 
%The work of David Vokrouhlick\'y was partly 
%supported by the Czech Grant Agency (grant 205/08/0064). 

\clearpage

\clearpage

\clearpage

\clearpage

\clearpage

\clearpage

\clearpage

\clearpage

\clearpage

\clearpage

\clearpage

\clearpage

\clearpage

\clearpage

\clearpage

\clearpage

\clearpage

\end{document}